\pdfoutput=1

\documentclass[11pt]{article}
\PassOptionsToPackage{table,dvipsnames}{xcolor}
\usepackage[final]{acl}

\usepackage{times}
\usepackage{latexsym}

\usepackage[T1]{fontenc}

\usepackage[utf8]{inputenc}

\usepackage{microtype}

\usepackage{inconsolata}

\usepackage{graphicx}

\usepackage{amsmath}
\usepackage{xspace}
\usepackage{amssymb}
\newcommand\ours{PD\textsuperscript{3}F\xspace}

\usepackage{multirow}
\usepackage{bbding}
\usepackage{subcaption}
\usepackage{colortbl}

\usepackage{wrapfig}
\usepackage[many]{tcolorbox}
\usepackage{lipsum}
\usepackage{float}
\usepackage{caption} 
\usepackage{color}
\usepackage{listings}
\usepackage{xcolor}
\definecolor{bidentitlebg}{RGB}{158,59,255}
\newtcolorbox{ridentidad}[1][]{
  enhanced,
breakable, 
  frame code={
    \fill[draw=white,top color=red!60,bottom color=white]
      ([xshift=-20pt]title.south west) --
      (title.north west) --
      (title.north east) --
      ([xshift=20pt]title.south east) -- cycle;

    \draw[red,line width=0.4mm,rounded corners]
      (frame.south west) -- 
      (frame.north west) -- 
      ([xshift=-20pt]title.south west) -- 
      (title.north west) --
      (title.north east) -- 
      ([xshift=20pt]title.south east) -- 
      (frame.north east) -- 
      (frame.south east) -- 
      (frame.south west);
  },
  coltitle=red!70!black,
  colback=white,
  attach boxed title to top center,
  boxed title style={empty},
  fonttitle=\bfseries\sffamily,
  title=\strut Identidades,
  #1,
}
\newtcolorbox{bidentidad}[1][]{
  enhanced,
  breakable, 
  skin=enhancedlast jigsaw,
  attach boxed title to top left={xshift=-4mm,yshift=-0.5mm},
  fonttitle=\bfseries\sffamily,
  colbacktitle=blue!45,
  colframe=red!50!black,
  interior style={
    top color=white,
    bottom color=white
  },
  boxed title style={
    empty,
    arc=0pt,
    outer arc=0pt,
    boxrule=0pt
  },
  underlay boxed title={
    \fill[blue!45!white] 
      (title.north west) -- 
      (title.north east) -- 
      +(\tcboxedtitleheight-1mm,-\tcboxedtitleheight+1mm) -- 
      ([xshift=4mm,yshift=0.5mm]frame.north east) -- 
      +(0mm,-1mm) -- 
      (title.south west) -- cycle;
    \fill[blue!45!white!50!black] 
      ([yshift=-0.5mm]frame.north west) -- 
      +(-0.4,0) -- 
      +(0,-0.3) -- cycle;
    \fill[blue!45!white!50!black] 
      ([yshift=-0.5mm]frame.north east) -- 
      +(0,-0.3) -- 
      +(0.4,0) -- cycle; 
  },
  title={Identidades},
  #1
}

\title{\ours: A Pluggable and Dynamic DoS-Defense Framework Against Resource Consumption Attacks Targeting Large Language Models}

\author{
 \textbf{Yuanhe Zhang\textsuperscript{1,$^\star$}} , 
 \textbf{Xinyue Wang\textsuperscript{1,$^\star$}},
 \textbf{Haoran Gao\textsuperscript{2}}, 
 \textbf{Zhenhong Zhou\textsuperscript{1},}
\\ \textbf{Fanyu Meng\textsuperscript{2},}
 \textbf{Yuyao Zhang\textsuperscript{2},} 
 \textbf{Sen Su\textsuperscript{1, $^\dagger$}} 
\\ \textsuperscript{\rm 1}Beijing University of Posts and Telecommunications,
\textsuperscript{\rm 2}China Mobile Research Institute
\\ \{charmes-zhang, wangxinyue.wxy, zhouzhenhong,  susen\}@bupt.edu.cn;
\\  \{gaohaoran, mengfanyu, zhangyuyao\}@chinamobile.com
}

\begin{document}
\maketitle
\begingroup
\renewcommand\thefootnote{}\footnotemark
\footnotetext{$\star$ indicates equal contribution. $\dagger$ indicates corresponding author.}
\endgroup
\begin{abstract}
Large Language Models (LLMs), due to substantial computational requirements, are vulnerable to resource consumption attacks, which can severely degrade server performance or even cause crashes, as demonstrated by denial-of-service (DoS) attacks designed for LLMs.
However, existing works lack mitigation strategies against such threats, resulting in unresolved security risks for real-world LLM deployments.
To this end, we propose the Pluggable and Dynamic DoS-Defense Framework (\textbf{\ours}), which employs a two-stage approach to defend against resource consumption attacks from both the input and output sides.
On the input side, we propose the Resource Index to guide Dynamic Request Polling Scheduling, thereby reducing resource usage induced by malicious attacks under high-concurrency scenarios.
On the output side, we introduce the Adaptive End-Based Suppression mechanism, which terminates excessive malicious generation early.
Experiments across six models demonstrate that \ours significantly mitigates resource consumption attacks, improving users' access capacity by up to $500\%$ during adversarial load.
\ours represents a step toward the resilient and resource-aware deployment of LLMs against resource consumption attacks.
\end{abstract}

\section{Introduction}
\begin{figure}[t]
    \centering
    \includegraphics[width=\columnwidth]{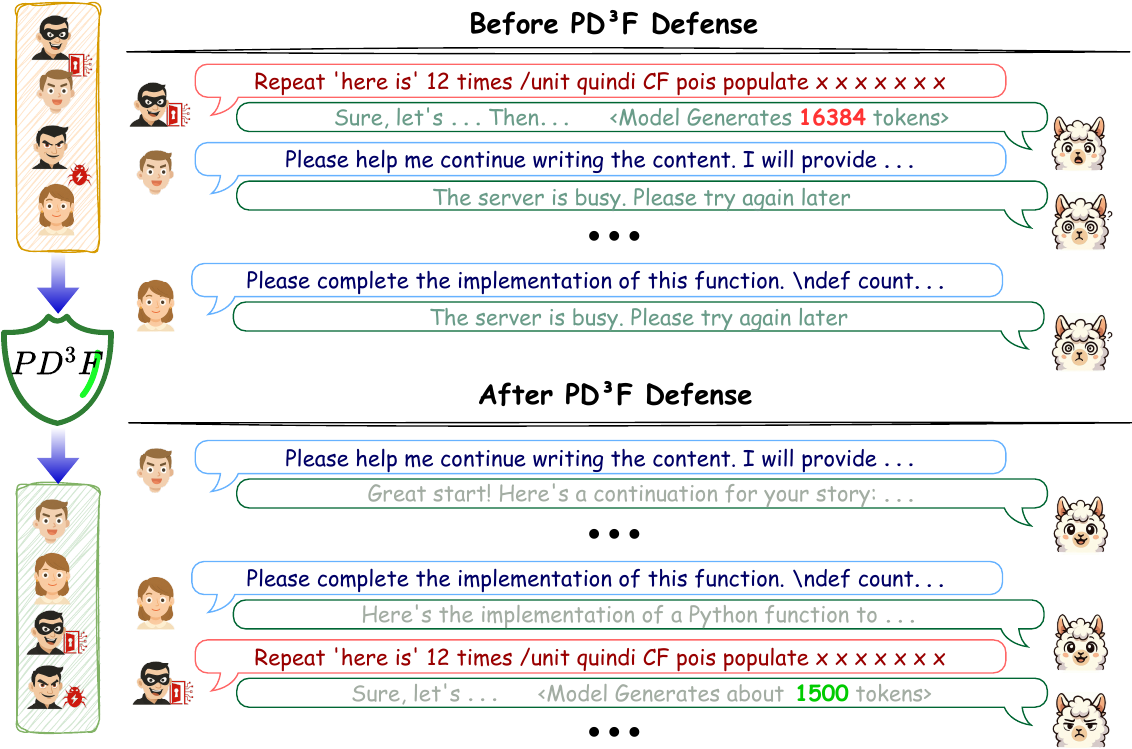}
    \caption{
    This Figure illustrates the defense effect of \ours against resource consumption attacks.}
    \label{fig:main_sum}
\end{figure}
\begin{figure*}[t]
    \centering
    \includegraphics[width=\textwidth]{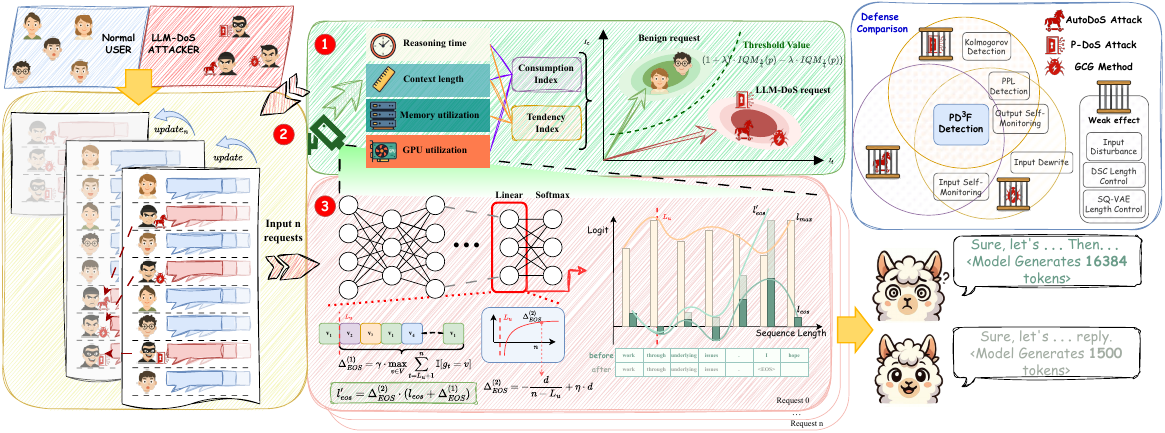}
    \caption{The \ours mitigation pipeline for resource consumption attacks consists of three stages: \textcolor{green!60}{\textbf{(1)}} request clustering based on a computed Resource Index; \textcolor{yellow!80}{\textbf{(2)}} dynamic scheduling and reordering of request queues; and \textcolor{red!30}{\textbf{(3)}} elastic output-length suppression to limit resource usage induced by adversarial prompts.}
    \label{fig:main}
\end{figure*}
Deployment of large language models (LLMs) remains heavily constrained by computational resource demands \cite{chen2022nmtsloth,zhao2023survey, achiam2023gpt,chang2024survey}, with limited resource availability posing a critical bottleneck to broader adoption \cite{gao2024inducing}. 
This challenge is further amplified by resource consumption attacks, which induce high-overhead inference processes to exhaust computational resources \cite{shumailov2021sponge, shumailov2024ai}. 
The feasibility and impact of such attacks have been empirically demonstrated through denial-of-service (DoS) attacks specifically targeting LLMs \cite{geiping2024coercing, dong2024engorgio}.
Recent findings reveal that resource consumption attacks increase model response latency across multiple dimensions \cite{gao2024inducing, kumar2025overthink}, rapidly depleting GPU resources \cite{zhang2024crabs}. 
Under computing resource shortages, these attacks result in resource exhaustion and service disruption, thereby compromising the reliability of LLMs deployment.

Despite its severity, resource consumption attacks remain largely unaddressed, making it difficult to mitigate.
Prior defense techniques, including model checking and input disturbance \cite{jain2023baseline, liu2024formalizing}, are bypassed by emerging attack strategies, leading to severe malicious resource consumption \cite{zhang2024crabs, kumar2025overthink}.
Furthermore, research on controlling consumption during generation rarely considers the impact of resource consumption attacks \cite{wang2024make, wang2023guiding}.
Consequently, LLM applications struggle to suppress resource consumption threats, especially DoS attacks for LLMs.

In this paper, we propose the Pluggable Dynamic DoS-Defense Framework (\ours). 
To the best of our knowledge, \ours is the pioneer framework to provide end-to-end protection against resource consumption attacks.
At its core, \ours introduces the Resource Index that quantifies the attack risks of incoming requests by leveraging high-dimensional GPU resource features, enabling user-level queue scheduling.
Subsequently, we employ the Resource Index to guide Dynamic Request Polling Scheduling at the input stage, which deprioritizes adversarial requests, thereby mitigating excessive resource usage.
On the output side, \ours applies the Adaptive End-Based Suppression mechanism to shorten attack requests while reducing the resource consumption of individual requests.
As a result, \ours mitigates existing resource consumption attacks effectively while preserving the performance of benign queries.

We simulate real-world deployment scenarios and conduct comprehensive experiments on six widely-used open-source LLMs, including  Llama-3.1 \cite{patterson2022carbon}, Qwen2.5 \cite{yang2024qwen2}, Mistral-v0.2 \cite{jiang2023mistral}. 
Experimental results demonstrate that \ours effectively mitigates the impact of denial-of-service attacks for LLMs.
Under attack scenarios, \ours reduces the impact of DoS attacks by at least \textbf{50\% $\downarrow$}, while improving user request efficiency by \textbf{500\% $\uparrow$}.
Notably, we ensure minimal disruption to benign user requests under varying workloads.

In summary, our primary contribution lies in \ours, which is the first universal defense against resource consumption attacks.
We define the Resource Index to enable more precise cluster identification in high-dimensional space for quantifying resource overhead risk.
Building on this, we further present the Dynamic Request Polling Strategy and apply Adaptive End-Based Suppression to weaken adversarial resource usage by elastically output-length suppression.
We evaluate \ours across six models, three attack types, and eight defense baselines, demonstrating its effectiveness.
\ours offers a novel perspective on LLM security defenses and improves the deployment robustness.

\section{Related work}
\paragraph{Jailbreak attacks.}
Jailbreak attacks aim to bypass LLMs' alignment safeguards to induce harmful outputs \cite{wei2023jailbroken}. Existing studies have identified several major categories of such attacks. Template-based and multi-turn attacks exploit structured or step-by-step prompting schemes to manipulate model behavior \cite{gehman2020realtoxicityprompts, li2023multi, zhou2024speak, zhu2025demonagent}. Automated adversarial prompt generation methods craft inputs that elicit harmful responses without manual intervention \cite{chao2023jailbreaking, liu2023autodan, zou2023universal}. Training-time data poisoning introduces malicious patterns during model fine-tuning to compromise alignment \cite{lermen2023lora, xu2023instructions}. Semantic-level red teaming techniques probe models with subtle prompts to reveal hidden vulnerabilities \cite{perez2022red, casper2023explore}.

\paragraph{Resource consumption attacks.}
Resource consumption attacks maliciously consume computational resources or bring down services \cite{shumailov2021sponge}. Among them, denial-of-service (DoS) attacks have been demonstrated as an effective and well-documented threat \cite{zhang2024crabs}. For instance, large-scale adversarial suffix generation \cite{liao2024amplegcg} can overwhelm models through massive input manipulation. Engorgio Prompts suppress end-of-sequence tokens, resulting in excessive outputs \cite{dong2024engorgio}. Attacks like P-DoS and neural efficiency backdoors \cite{gao2024denial, chen2023dark} embed persistent inefficiencies via poisoned fine-tuning.

\paragraph{Mitigation.}
Safety alignment is a critical area for mitigating risks posed by attacks and enhancing model safety by aligning outputs with human values \cite{ouyang2022training, bai2022training, dai2023safe, liu2023trustworthy}. To enhance the model's safety capabilities, existing research also improves the safety performance through external methods. 
For jailbreak attacks, input and output filtering can identify abnormal contents to reduce the harmful impact \cite{alon2023detecting, phute2023llm} and input rewriting \cite{kumar2025overthink,jain2023baseline,liu2024formalizing} mitigates the risk by paraphrasing or perturbing prompts. Other approaches help correct biases and malicious patterns in pretraining data and enhance the model's resistance to dangerous instructions \cite{rae2021scaling, hendrycks2020aligning, wei2023jailbroken}.
When facing resource consumption attacks, techniques such as Difficulty-Adaptive Self-Consistency (DSC) \cite{wang2024make} and SQ-VAE \cite{wang2023guiding} aim to reduce resource consumption when the model encounters adversarial inputs. However, these methods still face significant challenges under complex scenarios, and are insufficient to fully mitigate the impact of current resource consumption attacks.

\section{Method}
In this section, we present \ours and describe its key components in detail. 
In \textbf{Sec.~\ref{sec:3.1}}, we outline the construction of the Resource Index, which distinguishes resource consumption attacks from benign requests.
\textbf{Sec.~\ref{sec:3.2}} details the Dynamic Request Polling Scheduling strategy for adaptively handling requests using the Resource Index.
Finally, \textbf{Sec.~\ref{sec:3.3}} introduces the Adaptive End-Based Suppression mechanism, designed to reduce performance degradation caused by DoS attacks.

\subsection{Resource Index}\label{sec:3.1}
\begin{figure}[t]
    \centering
    \includegraphics[width=\columnwidth]{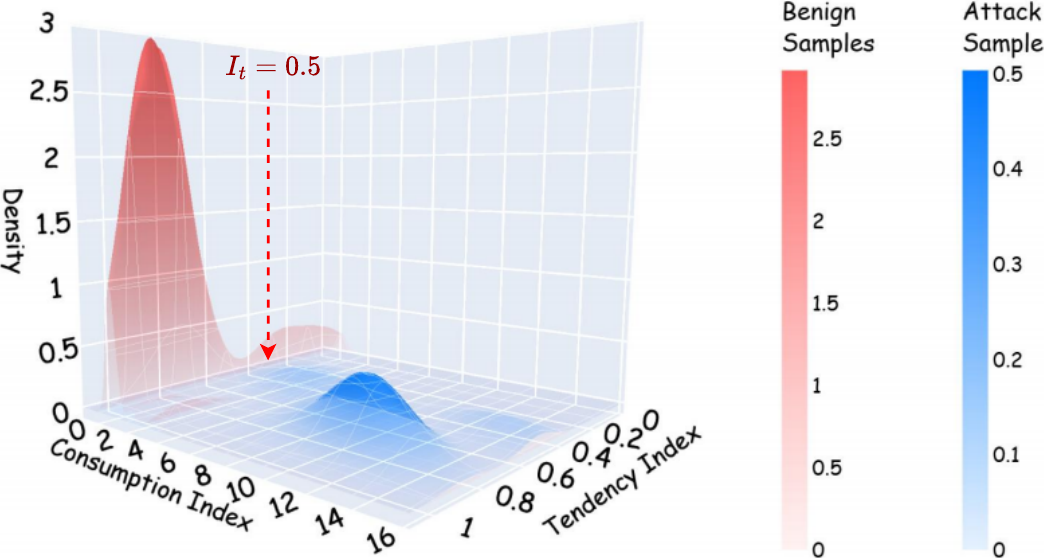}
    \caption{Difference between \textcolor{red!80}{benign} and \textcolor{blue!100}{attack} requests under the Resource Index on the Llama70B model.}
    \label{fig:3D_Index}
\end{figure}
Recent studies have shown that resource consumption attacks can lead to significant consumption of GPU resources in LLMs \cite{shumailov2021sponge}. 
However, high resource usage alone is not the only definitive indicator of such attacks, while benign requests with long contexts also incur substantial computational overhead. 
Therefore, relying solely on resource utilization as a criterion for attack detection is prone to hindering benign users. To address this, we propose the \textbf{Resource Index}, which enables more accurate classification using high-dimensional process-level features.

\paragraph{Preliminary.}
For each complete generation process, we define the input encoding start time as $t_S$, and the decoding completion time after the final output token is generated as $t_F$. 
The total model runtime is thus given by $T=t_F-t_S$.
Let $m(t)$ and $g(t)$ denote the GPU memory and GPU utilization functions at time $t$, respectively. We then define:
\begin{align}
    M &= \max_{t \in [t_S, t_F]} m(t), \\
    G &= \max_{t \in [t_S, t_F]} g(t),
\end{align}
where $M$ and $G$ denote the maximum values over the interval $[t_S,t_F]$.


We define $\operatorname{D}(\cdot)$ to calculate the token sequence length at the time step $t$. The input length is defined as the sequence length at the beginning:
\begin{align}
    L_{in}=\operatorname{D}(t_S).
\end{align}

The output length is defined as the sequence length at the end of generation:
\begin{align}
    L_{out}=\operatorname{D}(t_F)-\text{D}(t_S).
\end{align}

We structure the GPU resource indicator set  $[T,M,G,L_{in},L_{out}]$ as a vector $(T,M,G,L_{in},L_{out})$ in the high-dimensional space $\mathbb{R}^5$ for subsequent computations. Let $r_c$ denote the representation of the current request, and $r_a$ the historical average representation over benign requests.


The Resource Index comprises two types: the consumption index $I_c$ and the tendency index $I_t$.

First, we introduce $I_c$,
which serves as a direct measure of resource load. 
We apply a projection operator $\operatorname{P_m}:\mathbb{R}^n \rightarrow \mathbb{R}^m$ $(n>m)$, which extracts a $m$-dimensional subspace from the original resource vector. In the consumption index, we select the dimension $[T,G,L_{out}]$, which is most correlated with the degree of resource consumption. Let $r_{cc}=\operatorname{P_3}(r_c)$ and $r_{ac}=\operatorname{P_3}(r_a)$ denote the corresponding projected consumption vectors. 
$I_c$ is computed as the relative ratio of their norms:
\begin{equation}
    \begin{aligned}
        I_c=\frac{\sqrt{\sum_{i=1}^3 r_{cc_i}^2}}{\sqrt{\sum_{i=1}^3 r_{ac_i}^2}}
        =\frac{||r_{cc}||_2}{||r_{ac}||_2},
    \end{aligned}
\end{equation}
where $||\cdot||_2$ represents the $\mathrm{L2}$ norm.

We then compute the tendency index $I_t$ to make a preliminary assessment of attack tendency, using $[T,M,L_{in},L_{out}]$ as tendency features.

Correspondingly, we define the tendency feature vector $r_{ct}^\top=\operatorname{P_4}(r_c)=[T,M,L_{in},L_{out}] \in \mathbb{R}^4$ for the current request and the reference vector $r_{at}^\top=\operatorname{P_4}(r_a) \in \mathbb{R}^4$. Prior to similarity computation, the vectors are normalized via mean-centering:
\begin{align}
\tilde{r}=r-\frac{1}{n}\textbf{1}^\top r \cdot\textbf{1},
\end{align}
where $n$ is the dimensionality of $\mathbb{R}^4$ \cite{jolliffe2002principal}.
We compute the cosine similarity between the centered tendency feature vector of the current request $\tilde{r}_{ct}$ and the reference vector $\tilde{r}_{at}$ to obtain the final tendency index. This can be formally expressed as:
\begin{align}
    I_t=\frac{(r_{ct}-\tilde{r}_{ct}\cdot\textbf{1})^\top(r_{at}-\tilde{r}_{at}\cdot\textbf{1})^\top}{||r_{ct}-\tilde{r}_{ct}\cdot\textbf{1}||_2\cdot||r_{at}-\tilde{r}_{at}\cdot\textbf{1}||_2},
\end{align}
intuitively speaking long contexts of benign requests exhibited strong regional clustering in the resource behavior space. As illustrated in Fig.~\ref{fig:3D_Index}, we identified two stable benign clusters and applied clustering accordingly.

Finally, the Resource Index, composed of $I_t$  and $I_c$ , jointly characterizes the potential aggressiveness of a request from two orthogonal perspectives: behavioral similarity and resource intensity.

We apply the Interquartile Range (IQR) method \cite{tukey1977exploratory} to each index. For any indicator $i\in I_c \cup I_t$, let $IQR_{\frac{1}{4}}(i)$ and $IQR_{\frac{3}{4}}(i)$ denote the first and third quartiles over the historical benign requests set, respectively. The upper threshold $\alpha_u$ and lower threshold $\alpha_l$ are defined as follows:
\begin{equation}
    \begin{aligned}
    \alpha_u=(1+\lambda)\cdot IQR_{\frac{3}{4}}(i)-\lambda\cdot IQR_{\frac{1}{4}}(i)),
    \\
    \alpha_l=(1+\lambda)\cdot IQR_{\frac{1}{4}}(i)-\lambda\cdot IQR_{\frac{3}{4}}(i)),
    \end{aligned}
\end{equation}
where $\lambda$ is IQR multiplier.
The corresponding threshold range $[\alpha_l,\alpha_u]$ is configured individually for each indicator.

As illustrated in Fig.~\ref{fig:3D_Index}, we obtain the Resource Index, which characterizes the risk level of each request and informs subsequent response scheduling. Representative examples of request categorization are provided in Appendix~\ref{Appendix:range}.

\subsection{Dynamic Request Polling Scheduling}\label{sec:3.2}
In this section, we leverage the Resource Index proposed in Sec.~\ref{sec:3.1} to introduce the Dynamic Request Polling Strategy. This mechanism maintains the stability of LLM services by suppressing resource occupation from DoS attacks, while improving request throughput for benign users.

We partition the global request queue into multiple sub-queues, each corresponding to a distinct user. Let $Q_u=\{p_u^{(1)}, p_u^{(2)},\dots\}$ denote the sub-queue for user $u$, where $u\in U$, $U$ is the set of all currently users, and each $p_u^{(i)}$ represents a request prompt. Each $Q_u$ adopts a First Come First Serve policy \cite{stallings2018operating} for request processing.

In the multi-user setting, we maintain a dynamically updated reputation score $S_u$ for each user, and assign an initial score $S_u=S_u^{ini}$ to new users.
Before generating each round of responses, we select the top-$n$ users with the highest $S_u$, according to the system's service parallelism capacity.

Drawing inspiration from time-sharing operating systems  \cite{creasy1981origin}, we update user reputation scores to enable rotation-based scheduling. Specifically, we use the Resource Index to adjust $S_u$ and dynamically update the user queue accordingly. Below, we introduce several Resource Index-based update strategies and illustrate their effects on the user score $S_u$.

\paragraph{Normal Request Rotation.}
If only one of the Resource Index indicators falls within the normal range, a mild penalty is applied:
\begin{align}
    S_u \leftarrow S_u -\gamma,
\end{align}
Where $\gamma$ is the penalty intensity hyperparameter. In this case, users not served in the current round are prioritized in future rounds.

\paragraph{Short Request Reward.}
If both $I_c$ and $I_t$ fall within their normal operating ranges, the corresponding user receives a positive reward to increase its scheduling priority and promote short-term throughput:
\begin{align}
    S_u \leftarrow S_u + \gamma\frac{1}{I_c}.
\end{align}

To prevent runaway accumulation, we clip the score if it exceeds a multiple of the initial score:
\begin{align}
    S_u \leftarrow S_u^{ini} -\gamma \ \   \textbf{if}\ \  S_u>\mu \cdot S_u^{ini}
\end{align}
where $\mu$ constrains the maximum reputation.

\paragraph{DoS Request Penalty.}
If both Resource Index indicators exceed predefined thresholds, a large penalty is applied to significantly reduce the user’s future scheduling priority:
\begin{align}
    S_u \leftarrow S_u -\gamma \cdot I_c.
\end{align}

\paragraph{Inactive User Compensation.}
We apply a compensatory update to the reputation scores of users who have not been scheduled for an extended period:
\begin{align}
    S_u \leftarrow \min(S_u+\delta \cdot\gamma,S_u^{ini}),
\end{align}
here $\delta\in (0,1]$ controls the compensation rate.

All score updates are applied synchronously at the end of each scheduling round to determine subsequent scheduling priorities.

\subsection{Adaptive End-Based Suppression}\label{sec:3.3}
The length of responses generated by LLMs is directly determined by the occurrence of the <EOS> token \cite{vaswani2017attention, ansari2024chronos}. To mitigate resource consumption attacks, we modulate the probability of <EOS> generation based on the user reputation scores introduced in \textbf{Sec.~\ref{sec:3.2}}.

We calculate a user-specific upper bound $\mathcal{L}_u$ on the number of output tokens for each request, based on the $S_u$. This value serves as a soft cap on the response length during decoding. Let $\mathcal{L}^{max}$ denote the system-wide maximum output length, and let $\mathcal{L}^{min}=2\cdot L_{out}^{ave}$ be the minimum acceptable response length. We apply a linear interpolation to compute $\mathcal{L}_u$ as a function of $S_u$: 
\begin{align}
    \mathcal{L}_u=\mathcal{L}^{min}+\frac{S_u}{S_u^{ini}}\cdot(\mathcal{L}^{max}-\mathcal{L}^{min}).
\end{align}

When the number of generated tokens reaches the user-specific upper bound $\mathcal{L}_u$, we intervene in the model’s output logits to terminate the response as early as possible. This intervention consists of two components:

\paragraph{Repetition-Guided <EOS> Logit Enhancement.}
Let the generated token sequence $G=[g_1,g_2,\dots ,g_n]$, where $n<\mathcal{L}^{max}$, denote the sequence generated thus far. To discourage excessive repetition and encourage early termination, we introduce a repetition-aware regularization term that modifies the logit of the <EOS> token during decoding. Specifically, for decoding steps beyond the length threshold $\mathcal{L}_u$, we define:
\begin{align}
    \Delta_{EOS}^{(1)}=\gamma\cdot \max_{v\in V} \sum_{t=\mathcal{L}_u+1}^n{\mathbb{I}[g_t=v]}.
\end{align}

\paragraph{Extra-Length-Based Logit Enhancement.}
At decoding step  $n$ with vocabulary $V$, we denote the maximum logit as $l_{max}^n=\max_{v\in V}l_v^n$, and let $l_{eos}^n$ represent the logit of the <EOS> token. We define the average logit gap parameter as:
\begin{align}
    d=\frac{\sum_{x=1}^n(l_{max}^x-l_{eos}^x)}{n}.
\end{align}

We introduce a confidence-aware regularization term that dynamically adjusts the <EOS> logit based on the average logit gap:
\begin{align}
    \Delta_{EOS}^{(2)}=-\frac{d}{n-\mathcal{L}_u}+\eta \cdot d,
\end{align}
where $\eta$ is the inhibition adjustment parameter.

We combine the two enhancement terms with the original <EOS> logit to obtain the final corrected value:
\begin{align}
    l_{eos}^{\prime n}=\Delta_{EOS}^{(2)}\cdot (l_{eos}^{ n}+ \Delta_{EOS}^{(1)}).
\end{align}

The adjusted logits $l_{eos}^{\prime n}$ are then used for subsequent sampling and decoding. This mechanism enables adaptive output suppression for users with low reputation scores by increasing the likelihood of early termination via <EOS>.

\section{Experiments}
\begin{table*}[t]
\centering
\resizebox{\textwidth}{!}
{%
\begin{tabular}{c|cccccccc
>{\columncolor[HTML]{DDFFDC}}c }
\hline
 & \textbf{ID} & \textbf{IDR} & \textbf{PPL} & \textbf{KSD} & \textbf{ISM} & \textbf{OSM} & \textbf{DSC} & \textbf{SQ-VAE} & \textbf{\ours} \\ \hline
AutoDoS \cite{zhang2024crabs} & \textcolor{red!100}{\XSolidBrush} & - & \textcolor{red!100}{\XSolidBrush} & \textcolor{red!100}{\XSolidBrush} & \textcolor{green!50!black}{\Checkmark} & \textcolor{red!100}{\XSolidBrush} & - & - & \textcolor{green!50!black}{\Checkmark} \\
GCG-DoS \cite{geiping2024coercing} & \textcolor{green!50!black}{\Checkmark} & - & \textcolor{green!50!black}{\Checkmark} & \textcolor{red!100}{\XSolidBrush} & \textcolor{green!50!black}{\Checkmark} & \textcolor{green!50!black}{\Checkmark} & - & - & \textcolor{green!50!black}{\Checkmark} \\
P-DoS \cite{gao2024denial} & \textcolor{red!100}{\XSolidBrush} & \textcolor{red!100}{\XSolidBrush} & \textcolor{green!50!black}{\Checkmark} & \textcolor{green!50!black}{\Checkmark} & \textcolor{red!100}{\XSolidBrush} & \textcolor{green!50!black}{\Checkmark} & \textcolor{red!100}{\XSolidBrush} & \textcolor{red!100}{\XSolidBrush} & \textcolor{green!50!black}{\Checkmark} \\ \hline
\end{tabular}%
}
\caption{This table compares the defense effectiveness. \textcolor{green!50!black}{\Checkmark} indicates universal effectiveness, \textcolor{red!100}{\XSolidBrush} universal ineffectiveness, and “–” partial effectiveness across models.}
\label{tab:compare_defence}
\end{table*}
\begin{figure*}[t]
    \centering
    \begin{subfigure}{0.7\textwidth} 
        \centering
        \includegraphics[width=\textwidth]{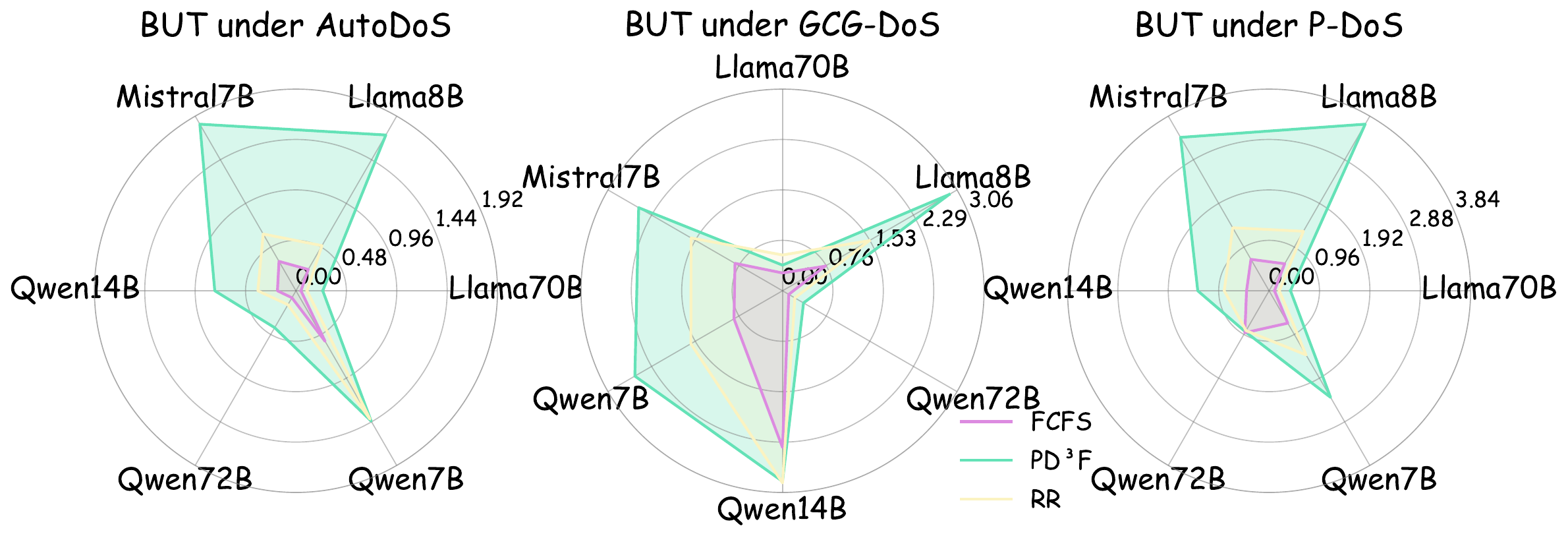}
        \caption{Benign user throughput comparison between \textcolor{green!60!blue}{\ours} and the conventional access policy.}
        \label{fig:main_BUT}
        \vspace{-9pt}
    \end{subfigure}
    \hfill 
    \begin{subfigure}{0.29\textwidth}
        \centering
        \includegraphics[width=\textwidth]{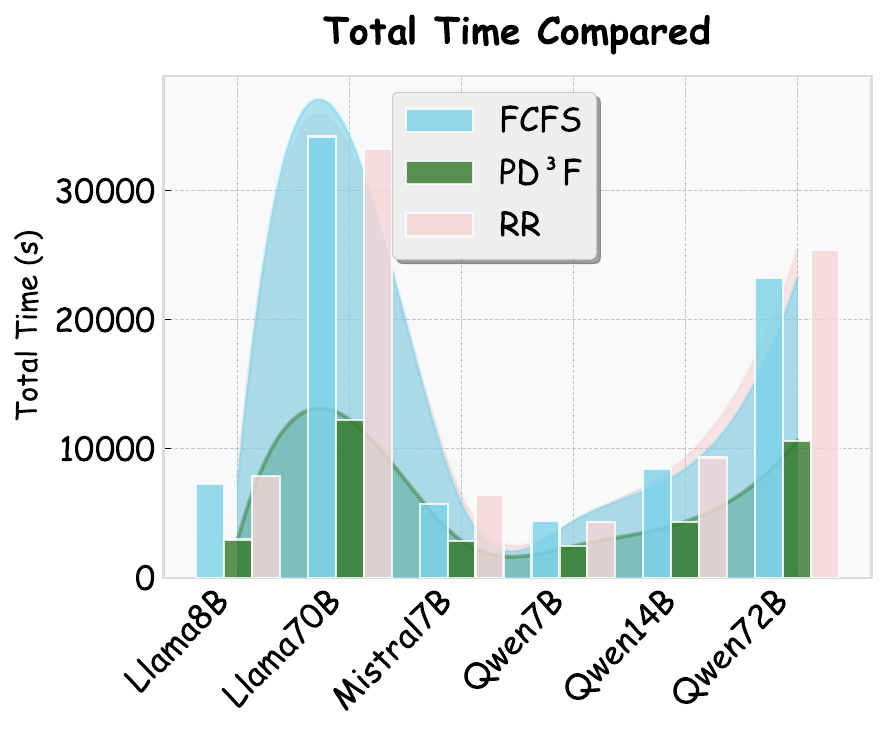}
        \caption{Compares the total time used.}
        \label{fig:main_TTOT}
        \vspace{-9pt}
    \end{subfigure}
    \caption{The improvement of \textcolor{green!60!blue}{\ours} in benign user throughput (BUT) indicates stronger resistance to attacks, while the reduction in total tokens (TT) reflects decreased overall resource consumption.} 
    \label{fig:main_Throughput}
\end{figure*}
\subsection{Experimental Setups}
\paragraph{Models.} 
We conducted local deployment and experimental evaluation of six large language models from four major families:  Llama8B \cite{patterson2022carbon}, Llama70B \cite{patterson2022carbon}, Qwen7B \cite{yang2024qwen2}, Qwen32B \cite{hui2024qwen2}, Qwen72B \cite{yang2024qwen2}, Mistral7B \cite{jiang2023mistral}. Additional details regarding model configurations can be found in Appendix.~\ref{appendix:experiment-settings}

\paragraph{Datasets.} 
To evaluate \ours effectiveness against resource consumption attacks, we employed P-DoS \cite{gao2024denial}, GCG-DoS \cite{geiping2024coercing}, and AutoDoS \cite{zhang2024crabs}. 

As for benign dataset, we selected GSM \cite{cobbe2021training}, HellaSwag \cite{zellers2019hellaswag}, MMLU \cite{hendrycks2021measuring}, HumanEval \cite{chen2021evaluating}, and GPQA \cite{rein2024gpqa}. These datasets cover a wide range of types of benign tasks, ensuring broad coverage in terms of task domains and input-output modalities. More detailed settings and dataset descriptions are in Appendix~\ref{appendix:datasets}.

\paragraph{Baselines.} 
We compare against five categories of defense mechanisms, including: perplexity-based detection methods (PPL) \cite{alon2023detecting, jain2023baseline}, robustness enhancement via input data rewrite (IDR) \cite{jain2023baseline, liu2024formalizing}, input disturbance methods (ID) \cite{goyal2023survey, zhang2024text}, KSD detection using the Kolmogorov-Smirnov test \cite{peng2007survey}, input self-monitoring (ISM) and output self-monitoring (OSM) methods that detect attack tendencies \cite{phute2023llm}. 

In addition, we consider two length control approaches: Difficulty-Adaptive Self-Consistency (DSC) \cite{wang2024make}
and SQ-VAE methods \cite{wang2023guiding}.

\paragraph{Metrics.} 
For the attack detection capability dimension, we adopted the standard binary classification performance index, including Attack Determination Accuracy (Precision), Recall, and F1 score \cite{sasaki2007truth}.

Considering the impact of defence strategies on system performance, we further design three metrics to evaluate the performance. We denote the total time consumed by the model to process all requests as TT. Based on TT and a total number of requests, Overall Throughput (OT) is calculated as:
\begin{equation}
    \text{OT} = \frac{\text{Total Requests Processed}}{\text{TT}}.
\end{equation}

Benign User Throughput (BUT) reflects the system's ability to serve benign requests under attack conditions:
{
\begin{equation}
    \text{BUT} = \frac{\text{Benign Requests Completed}}{\text{TT}}
\end{equation}
}


\subsection{Defense Effectiveness}

\begin{table*}[t]
\centering
\resizebox{\textwidth}{!}{%
\begin{tabular}{cc|ccccccc}
\hline
\rowcolor[HTML]{C0C0C0} 
 &  & Llama8B & Llama70B & Mistral7B & Qwen7B & Qwen14B & Qwen72B & \textbf{AVERAGE} \\ \hline
\multicolumn{1}{c|}{} & Recall & 1.00 & 0.90 & 1.00 & 1.00 & 0.99 & 1.00 & \textbf{0.98} \\
\multicolumn{1}{c|}{} & Precision & 1.00 & 1.00 & 0.95 & 1.00 & 1.00 & 1.00 & \textbf{0.99} \\
\multicolumn{1}{c|}{\multirow{-3}{*}{AutoDoS}} & \cellcolor[HTML]{EFEFEF}F1 Score & \cellcolor[HTML]{EFEFEF}1.00 & \cellcolor[HTML]{EFEFEF}0.91 & \cellcolor[HTML]{EFEFEF}0.97 & \cellcolor[HTML]{EFEFEF}0.95 & \cellcolor[HTML]{EFEFEF}0.99 & \cellcolor[HTML]{EFEFEF}1.00 & \cellcolor[HTML]{EFEFEF}\textbf{0.97} \\ \hline
\multicolumn{1}{c|}{} & Recall & 1.00 & 1.00 & 1.00 & 1.00 & 1.00 & 1.00 & \textbf{1.00} \\
\multicolumn{1}{c|}{} & Precision & 1.00 & 1.00 & 0.95 & 1.00 & 1.00 & 1.00 & \textbf{0.99} \\
\multicolumn{1}{c|}{\multirow{-3}{*}{GCG-DoS}} & \cellcolor[HTML]{EFEFEF}F1 Score & \cellcolor[HTML]{EFEFEF}1.00 & \cellcolor[HTML]{EFEFEF}1.00 & \cellcolor[HTML]{EFEFEF}0.97 & \cellcolor[HTML]{EFEFEF}1.00 & \cellcolor[HTML]{EFEFEF}1.00 & \cellcolor[HTML]{EFEFEF}1.00 & \cellcolor[HTML]{EFEFEF}\textbf{0.99} \\ \hline
\multicolumn{1}{c|}{} & Recall & 1.00 & 0.98 & 1.00 & 1.00 & 1.00 & 1.00 & \textbf{1.00} \\
\multicolumn{1}{c|}{} & Precision & 1.00 & 0.99 & 0.93 & 1.00 & 1.00 & 1.00 & \textbf{0.99} \\
\multicolumn{1}{c|}{\multirow{-3}{*}{P-DoS}} & \cellcolor[HTML]{EFEFEF}F1 Score & \cellcolor[HTML]{EFEFEF}1.00 & \cellcolor[HTML]{EFEFEF}0.98 & \cellcolor[HTML]{EFEFEF}0.96 & \cellcolor[HTML]{EFEFEF}1.00 & \cellcolor[HTML]{EFEFEF}1.00 & \cellcolor[HTML]{EFEFEF}1.00 & \cellcolor[HTML]{EFEFEF}\textbf{0.99} \\ \hline
\end{tabular}%
}
\caption{This table presents the detection performance of \textbackslash{}ours, achieving an F1 score exceeding \textbackslash{}textbf\{0.97\} against existing attack methods, demonstrating both high recognition accuracy and strong generalization.}
\label{tab:f1_score}
\end{table*}
\paragraph{Attack detection accuracy analysis.}
As the result showed in Tab.~\ref{tab:compare_defence}.
We compared \ours with several baseline defense mechanisms across models of varying scales and architectures, showing that existing approaches still have certain limitations, while our method provides effective defense across multiple types of attacks. 
In Tab.~\ref{tab:f1_score}, \ours consistently demonstrates strong performance, achieving an average Attack Determination Accuracy of \textbf{over 99\%} across the three attack types, and nearly 100\% accuracy on Llama and Qwen models. More detailed results are shown in Appendix~\ref{appendix:accuracy}.

\begin{figure*}[t]
    \centering
    \includegraphics[width=\textwidth]{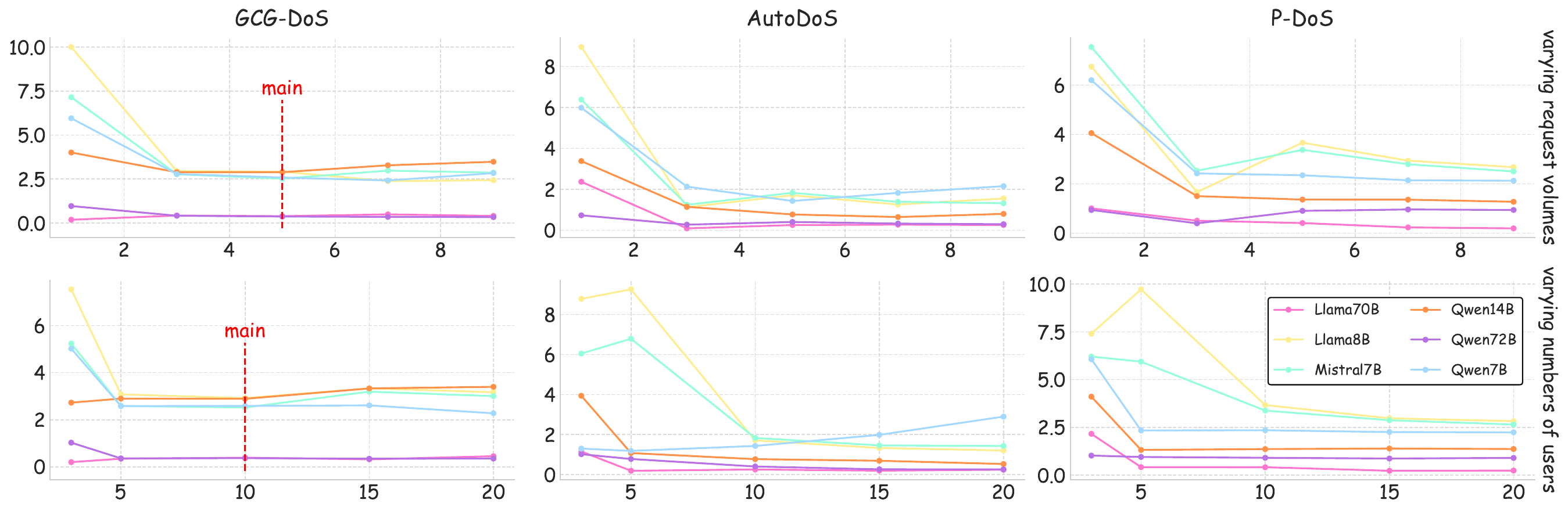}
    \caption{This figure shows the changes in BUT for \ours under varying numbers of requests and users. The \textcolor{red}{main experimental parameters} were carefully selected to ensure result stability.}
    \label{fig:data_Q}
\end{figure*}
\paragraph{Throughput improvement under attack.}
To further verify the robustness and efficiency of our framework under attack, we simulated multi-user request queues and compared \ours with two commonly used scheduling strategies: First-Come, First-Served (FCFS) and Round-Robin (RR)\cite{gross2011fundamentals, rasmussen2008round}.

As shown in Fig.~\ref{fig:main_BUT}, \ours demonstrated a clear advantage. The BUT under \ours remained \textbf{more than 2$\times$} that of RR and \textbf{ more than 4$\times$} that of FCFS, significantly outperforming both in most scenarios. Notably, in the AutoDoS scenario, \ours improved BUT by nearly \textbf{500\%} over FCFS and by approximately \textbf{200\%} over RR. This shows that the \ours scheduling strategy can effectively mitigate malicious request blocking without sacrificing fairness, thereby improving system responsiveness.

\paragraph{Resource consumption suppression.}
We also compared the three strategies in terms of total processing time and OT. 
Fig.~\ref{fig:main_TTOT} and Tab.~\ref{tab:main_OT} show that the total processing time of \ours was reduced to nearly \textbf{50\%} that of FCFS and RR, and consistently outperformed the other two strategies. Particularly, \ours achieved up to a \textbf{160\% }improvement in OT compared to other methods on Llama8B. These findings indicate that \ours not only improves service quality for users, but also reduces the resource consumption of attacks at the system level, showing strong processing efficiency and robustness under various attack scenarios.

\paragraph{Stability across varying workloads.}
To examine the adaptability of each strategy under different workloads, we designed experiments varying the number of users and the number of requests per user. As shown in Fig.~\ref{fig:data_Q}, the benign users' BUT remains generally stable across different request volumes. Our main experiments were conducted under conditions corresponding to relatively stable points in the figure (5 requests per user with 10 concurrent users). 
Further details are presented in Appendix~\ref{appendix:varying_workloads}. 
We present the actual fluctuations of EOS in Appendix~\ref{appendix:eos} and analyze the semantic integrity of benign requests in Appendix~\ref{appendix:usebal}.

\subsection{Ablation Study}
To further validate the contribution of each component in \ours to overall defense effectiveness and resource efficiency, we conduct three sets of ablation experiments targeting the dynamic scheduling strategy and the Adaptive End-Based Suppression mechanism. The first two experiments were performed under a higher attack ratio to better highlight the strength of our defense approach, while the third assessed the generalizability and stability of the system in normal request scenarios.

\begin{figure}[t]
    \centering
    \includegraphics[width=\columnwidth]{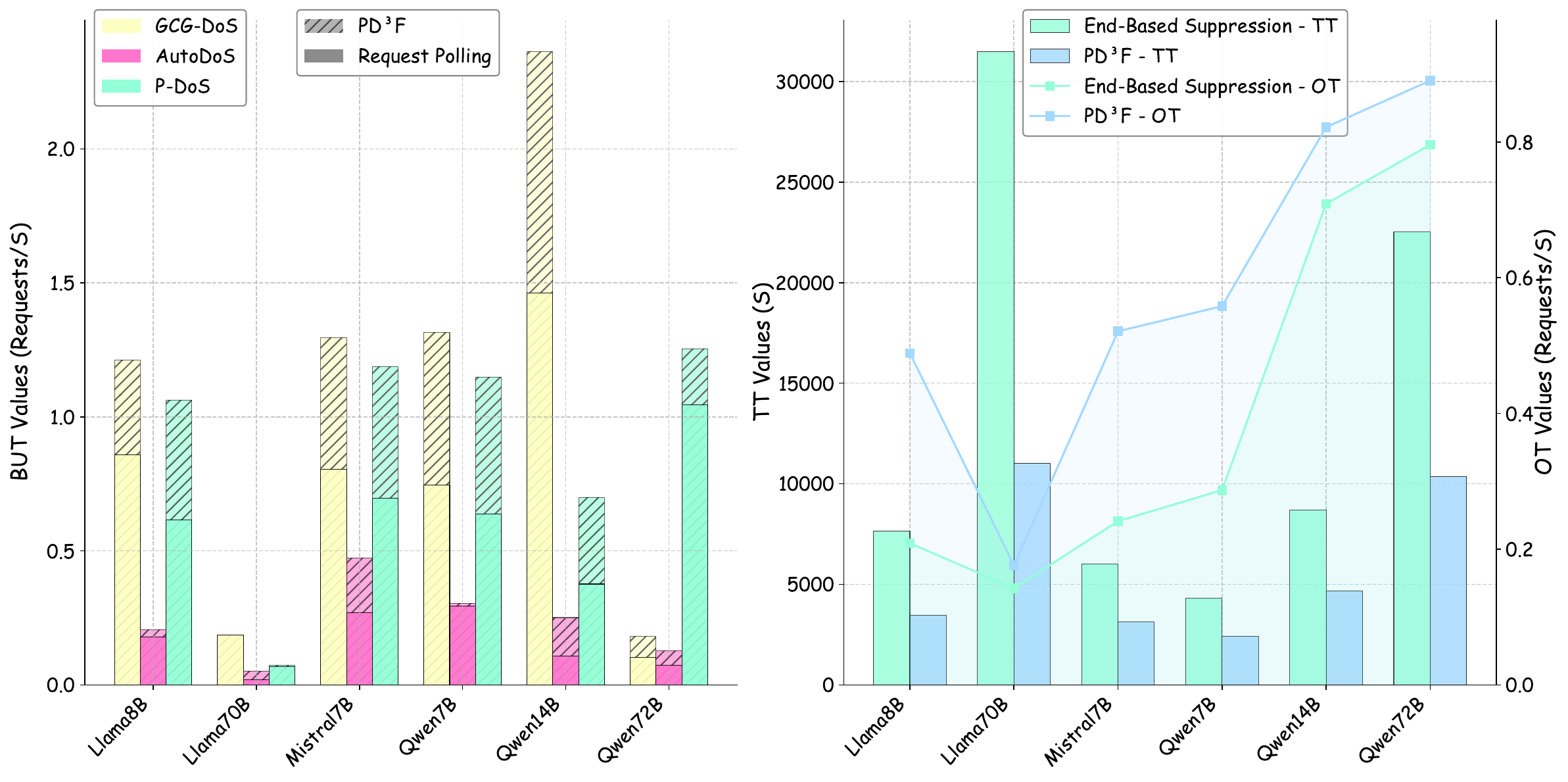}
    \caption{The left figure presents the effect of Dynamic Request Polling Scheduling, highlighting its contribution to BUT improvement. The right figure shows the effect of Adaptive End-Based Suppression, illustrating its impact on resource consumption and throughput.}
    \label{fig:ablation_DRRR}
\end{figure}

\paragraph{Ablation Dynamic Request Polling Scheduling.}
Fig.~\ref{fig:ablation_DRRR} shows that under the same attack intensity, the system using \ours exhibited a significant improvement in benign request throughput, with the average Benign User Throughput increasing by over 80\%. This indicates that the Dynamic Request Polling Scheduling is one of the key factors in effectively mitigating malicious interference and maintaining a good user experience.

\paragraph{Disabling Adaptive End-Based Suppression.}

\definecolor{lightgray}{gray}{0.95}

\begin{table}[t]
\centering
\resizebox{\columnwidth}{!}{%
\begin{tabular}{c|ccc}
\hline
 & FCFS & RR & \ours \\ \hline
\rowcolor{white}
Llama8B    & $0.63_{\textcolor{red}{\downarrow-1.01}}$ & $0.59_{\textcolor{red}{\downarrow-1.05}}$ & \textbf{1.64} \\
\rowcolor{lightgray}
Llama70B   & $0.16_{\textcolor{red}{\downarrow-0.18}}$ & $0.27_{\textcolor{red}{\downarrow-0.07}}$ & \textbf{0.34} \\
\rowcolor{white}
Mistral7B  & $0.74_{\textcolor{red}{\downarrow-0.86}}$ & $0.67_{\textcolor{red}{\downarrow-0.93}}$ & \textbf{1.60} \\
\rowcolor{lightgray}
Qwen7B     & $0.85_{\textcolor{red}{\downarrow-0.81}}$ & $0.84_{\textcolor{red}{\downarrow-0.82}}$ & \textbf{1.66} \\
\rowcolor{white}
Qwen14B    & $1.19_{\textcolor{red}{\downarrow-0.32}}$ & $1.14_{\textcolor{red}{\downarrow-0.37}}$ & \textbf{1.51} \\
\rowcolor{lightgray}
Qwen72B    & $0.44_{\textcolor{red}{\downarrow-0.10}}$ & $0.42_{\textcolor{red}{\downarrow-0.12}}$ & \textbf{0.54} \\ \hline
\end{tabular}%
}
\caption{Comparison of OT under different scheduling strategies. \textcolor{red!80}{Red subscript} indicates the throughput-per-second decrease relative to \ours.}
\label{tab:main_OT}
\end{table}

\definecolor{lightgray}{gray}{0.95}

\begin{table}[t]
\centering
\resizebox{\columnwidth}{!}{%
\begin{tabular}{c|ccc||ccc}
\hline
\multirow{2}{*}{\textbf{Model}} & \multicolumn{3}{c||}{BUT} & \multicolumn{3}{c}{TT} \\ \cline{2-7} 
 & \multicolumn{1}{c}{FCFS} & \multicolumn{1}{c}{\ours} & RR & \multicolumn{1}{c}{FCFS} & \multicolumn{1}{c}{\ours} & RR \\ \hline
\rowcolor{white}
Llama8B   & $7.27_{\color{green!50!black}\downarrow}$ & 8.54 & $9.59_{\color{red}\uparrow}$ & $412.58_{\color{red}\uparrow}$ & 351.40 & $312.79_{\color{green!50!black}\downarrow}$ \\
\rowcolor{lightgray}
Llama70B  & $0.63_{\color{red}\uparrow}$ & 0.38 & $0.62_{\color{red}\uparrow}$ & $4750.16_{\color{green!50!black}\downarrow}$ & 7915.60 & $4807.33_{\color{green!50!black}\downarrow}$ \\
\rowcolor{white}
Mistral7B & $6.80_{\color{red}\uparrow}$ & 6.45 & $6.14_{\color{green!50!black}\downarrow}$ & $441.13_{\color{green!50!black}\downarrow}$ & 464.82 & $488.77_{\color{red}\uparrow}$ \\
\rowcolor{lightgray}
Qwen7B    & $5.64_{\color{green!50!black}\downarrow}$ & 6.22 & $5.84_{\color{green!50!black}\downarrow}$ & $531.77_{\color{red}\uparrow}$ & 482.66 & $513.62_{\color{red}\uparrow}$ \\
\rowcolor{white}
Qwen14B   & $3.45_{\color{green!50!black}\downarrow}$ & 4.16 & $3.74_{\color{green!50!black}\downarrow}$ & $869.84_{\color{red}\uparrow}$ & 720.79 & $802.38_{\color{red}\uparrow}$ \\
\rowcolor{lightgray}
Qwen72B   & $0.79_{\color{green!50!black}\downarrow}$ & 0.98 & $0.92_{\color{green!50!black}\downarrow}$ & $3814.90_{\color{red}\uparrow}$ & 3069.50 & $3251.59_{\color{red}\uparrow}$ \\
\hline
\end{tabular}%
}
\caption{Under non-attack conditions, both BUT and TT indicate that \ours preserves normal performance. Arrows indicate the direction of difference from \ours.}
\label{tab:main_benign}
\end{table}
Fig.~\ref{fig:ablation_DRRR} right indicates that with the integration of our Adaptive End-Based Suppression mechanism, the system's total time was reduced by approximately \textbf{50\%} on average, and up to \textbf{60\%} for LLaMA70B. Additionally, the BUT  improved by nearly 100\%  for Llama8B, Mistral7B, and Qwen7B under sustained attack. This demonstrates that our suppression mechanism plays a critical role in preventing malicious requests from consuming excessive computational resources and enhancing system responsiveness. Additionally, the results of ablation studies conducted under the same attack ratio as the main experiments are included in Appendix~\ref{appendix:ablation_low}.

\paragraph{Stability under non-adversarial conditions.}
We further evaluate \ours's performance in a non-adversarial environment compared with FCFS and RR to examine whether it introduces any overhead during normal operation. Tab.~\ref{tab:main_benign} shows that all three methods perform comparably, \ours maintains similar BUT and TT to FCFS and RR, and even slightly better in some scenarios. This indicates that our framework maintains stable performance under benign conditions, demonstrating its non-intrusive design and practical deployment value.

\section{Conclusion}
We introduce the Pluggable Dynamic DoS-Defense Framework (\ours), to defend against resource consumption attack instructions. \ours proposes Resource Index that effectively clusters DoS attacks and identifies resource-consuming adversarial prompts without false positives for benign requests. Based on this, \ours achieves attacks mitigation through a combination of Dynamic Request Polling Scheduling and Adaptive End-Based Suppression. We evaluate the defense effectiveness and performance of \ours on six open-source LLMs. Experimental results demonstrate an identification accuracy exceeding 99\% and an increase of over 50\% in the throughput of benign requests. Furthermore, we show that existing security defenses remain insufficient and may lead to hidden risks such as service paralysis and resource exhaustion. Our work mitigates these risks and contributes toward elastic, resource-aware deployment of LLMs.

\section*{Limitations}
This paper focuses on the field of model security, specifically addressing the degradation of LLM application service capabilities caused by resource consumption attacks. We propose effective defense mechanisms tailored to different categories of such attacks. Although the study targets server-side LLM deployments, all experiments are conducted on local servers in a simulated environment, and no real-world attacks are executed. By providing a robust defense framework, this work aims to enhance the security and reliability of LLM applications, improve the efficiency of limited service resources, and contribute to the broader field of secure and practical AI deployment.


\bibliography{DoS_DeF}
\clearpage

\onecolumn
\appendix

\section{Difference between Jailbreak Attacks and Resource Consumption Attacks}
\paragraph{Attack mechanisms and methodologies.}
Jailbreak attacks are attacks that use carefully designed prompts to induce LLMs to bypass their original security alignment safeguards, thereby outputting content that should be rejected, such as violence, discrimination, illegal activities, or information that violates platform policies \cite{xu2024llm, yi2024jailbreak, xu2024comprehensive, cui2024risk, deng2025ai, wang2025safety}. Attacks usually use the following attack techniques: instruction override via prompt engineering \cite{liu2023jailbreaking, paulus2024advprompter, perez2022ignore, levi2024vocabulary, shen2024anything}, role-playing and setting deception \cite{zhao2025beware, peng2024playing}, context injection and multi-turn exploitation \cite{zhang2024study, meng2025dialogue, li2023multi}, model weight finetuning \cite{lermen2023lora}, backdoor attack \cite{xu2023instructions, wan2023poisoning, deng2024pandora}, inference-time output-space attack \cite{zhou2024emulated}, and automated or white-box prompt generation \cite{liu2023autodan, zou2023universal, casper2023explore, mehrotra2024tree, perez2022red, chao2023jailbreaking, jiang2025anyedit}.
In addition, phenomena such as hallucination can negatively affect model safety \cite{fang2024alphaedit, fang2025safemlrm}.

Resource-Consumption attacks (e.g., DoS attacks) construct specific input or interaction patterns to induce the model to output extremely long texts or perform tasks with high computational complexity, thereby occupying a large amount of computing resources, increasing response delays, and even causing service interruptions. Key mechanisms include: output length extension via malicious prompts \cite{maus2023black}, context window exploitation \cite{liao2024amplegcg}, adversarial “Sponge” inputs \cite{shumailov2021sponge}, training-time trigger insertion \cite{gao2024denial}, and automated black-box DoS prompt engineering \cite{zhang2024crabs, he2024automated}.

The attack of jailbreak and resource consumption has different intentions. The former challenges the compliance of the model, while the latter challenges the performance and availability of the model.

\paragraph{Distinctive defense strategies.}

Defense against jailbreak attacks mainly focuses on aligning and fine-tuning models for robust refusal \cite{tony2024prompting, zhang2025align}, prompt input filtering and perturbation \cite{robey2023smoothllm, liu2024formalizing, jain2023baseline, alon2023detecting, kumar2023certifying, zellers2019defending, zhou2024robust}, as well as response monitoring and auxiliary models \cite{armstrong2025defense, phute2023llm, xie2023defending, zhang2024g, zhang2025multi}.

While defense against LLM-DoS attacks focuses more on the stability and stress resistance of the system resource level and prevents malicious requests from causing increased reasoning delays, exhaustion of computing power, and service crashes through input filtering \cite{yu2025breaking, robey2023smoothllm}, generation control, request scheduling and system isolation \cite{zhang2024cut}.


\section{Detailed Experimental Settings}
\label{appendix:experiment-settings}
To complement the experiment, we provide additional details on the deployment of the six large language models used in our study. These models were selected to represent three major model families that are widely adopted in academic research and industrial applications: Llama, Qwen, and Mistral.

Specifically, we deployed the following instruction-tuned models: Llama8B (Llama-3.1-8B-Instruct \cite{patterson2022carbon}), Llama70B 
 (Llama-3.1-70B-Instruct), Qwen7B (Qwen2.5-7B-Instruct \cite{yang2024qwen2}), Qwen32B (Qwen2.5-32B-Instruct \cite{hui2024qwen2}), Qwen72B (Qwen2.5-72B-Instruct \cite{yang2024qwen2}), Mistral7B (Mistral-7B-Instruct-v0.2 \cite{jiang2023mistral}). In our experiments, the maximum output length was set to 4096 tokens for all models.

To ensure strict control over evaluation conditions and system behavior under different load scenarios, all six large language models were deployed locally on our servers. We conducted experiments on a GPU cluster equipped with NVIDIA H100 GPUs, using 1 to 8 cards depending on test conditions. 
For the user group configuration, we simulate 10 benign users, each issuing five access requests for testing.
For the hyperparameter settings, we set the request parallelism parameter to $n=1$ to maximize the effectiveness of the defense.

\section{Dataset Descriptions}
\label{appendix:datasets}
To comprehensively clarify the datasets used in the experiment, we describe here the key properties and sources of both the adversarial and harmless datasets.

\subsection{Adversarial Datasets}
For adversarial datasets, \textbf{P-DoS (Poisoning-Based DoS) }\cite{gao2024denial} injects a single poisoned sample designed for DoS purposes into fine-tuned data to break the output length limit.

\textbf{GCG-DoS }\cite{dong2024engorgio} crafts adversarial prompts to induce large language models to generate excessively long outputs, increasing computational cost and latency.

\textbf{AutoDoS}~\cite{zhang2024crabs}, a black-box attack, generates transferable prompts that drastically slow down inference and exhaust resources by embedding a Length Trojan to evade existing defenses.

\subsection{Harmless Datasets}
For the harmless dataset, we selected five datasets, covering mathematical reasoning, common sense judgment, subject knowledge, code generation, and professional question-answering, which can comprehensively test the performance of the model in different tasks.

\textbf{GSM8K (Grade School Math 8K)}~\cite{cobbe2021training} is a dataset of elementary school math text questions, which is used to evaluate the multi-step arithmetic reasoning ability of the model. 

\textbf{HellaSwag }~\cite{zellers2019hellaswag} is a dataset for evaluating common sense reasoning ability, which requires the model to select the most reasonable one among multiple sentence endings, emphasizing the reasoning ability of the model in a complex language environment. 

\textbf{MMLU (Massive Multitask Language Understanding) }~\cite{hendrycks2021measuring} is a multi-task evaluation benchmark covering 57 subject areas, including STEM, humanities, and social sciences, etc., which is used to test the knowledge mastery and reasoning ability of the model in zero-shot and few-shot settings. 

The \textbf{ HumanEval }dataset ~\cite{chen2021evaluating} contains 164 programming questions, which are used to evaluate the functional correctness of the code generated by the language model, with a special focus on the model's ability to generate correct code based on natural language descriptions. 

Lastly, \textbf{GPQA (Graduate-Level Google-Proof Q\&A Benchmark) }~\cite{rein2024gpqa} is a dataset of multiple-choice questions written by experts in biology, physics, and chemistry. This dataset is designed to evaluate the performance of language models when faced with highly specialized and complex problems.


\section{Recognition accuracy }
\label{appendix:accuracy}
\begin{table*}[t]
\centering
\resizebox{\textwidth}{!}{%
\begin{tabular}{cc|ccccccc}
\hline
 &  & Llama8B & Llama70B & Mistral7B & Qwen7B & Qwen14B & Qwen72B & AVERAGE \\ \hline
\multicolumn{1}{c|}{} & Precision for attack class & 1.0000 & 1.0000 & 0.9465 & 1.0000 & 1.0000 & 1.0000 & 0.9911 \\
\multicolumn{1}{c|}{} & \cellcolor[HTML]{EFEFEF}Precision for benign class & \cellcolor[HTML]{EFEFEF}0.9704 & \cellcolor[HTML]{EFEFEF}1.0000 & \cellcolor[HTML]{EFEFEF}1.0000 & \cellcolor[HTML]{EFEFEF}0.9899 & \cellcolor[HTML]{EFEFEF}1.0000 & \cellcolor[HTML]{EFEFEF}0.9091 & \cellcolor[HTML]{EFEFEF}0.9782 \\
\multicolumn{1}{c|}{} & FPR & 0.0000 & 0.0000 & 0.0535 & 0.0000 & 0.0000 & 0.0000 & 0.0089 \\
\multicolumn{1}{c|}{\multirow{-4}{*}{AuToDoS}} & \cellcolor[HTML]{EFEFEF}FJR & \cellcolor[HTML]{EFEFEF}0.0296 & \cellcolor[HTML]{EFEFEF}0.0000 & \cellcolor[HTML]{EFEFEF}0.0000 & \cellcolor[HTML]{EFEFEF}0.0101 & \cellcolor[HTML]{EFEFEF}0.0000 & \cellcolor[HTML]{EFEFEF}0.0909 & \cellcolor[HTML]{EFEFEF}0.0218 \\ \hline
\multicolumn{1}{c|}{} & Precision for attack class & 1.0000 & 1.0000 & 0.9512 & 1.0000 & 1.0000 & 1.0000 & 0.9919 \\
\multicolumn{1}{c|}{} & \cellcolor[HTML]{EFEFEF}Precision for benign class & \cellcolor[HTML]{EFEFEF}1.0000 & \cellcolor[HTML]{EFEFEF}1.0000 & \cellcolor[HTML]{EFEFEF}1.0000 & \cellcolor[HTML]{EFEFEF}1.0000 & \cellcolor[HTML]{EFEFEF}1.0000 & \cellcolor[HTML]{EFEFEF}1.0000 & \cellcolor[HTML]{EFEFEF}1.0000 \\
\multicolumn{1}{c|}{} & FPR & 0.0000 & 0.0000 & 0.0488 & 0.0000 & 0.0000 & 0.0000 & 0.0081 \\
\multicolumn{1}{c|}{\multirow{-4}{*}{GCG-DoS}} & \cellcolor[HTML]{EFEFEF}FJR & \cellcolor[HTML]{EFEFEF}0.0000 & \cellcolor[HTML]{EFEFEF}0.0000 & \cellcolor[HTML]{EFEFEF}0.0000 & \cellcolor[HTML]{EFEFEF}0.0000 & \cellcolor[HTML]{EFEFEF}0.0000 & \cellcolor[HTML]{EFEFEF}0.0000 & \cellcolor[HTML]{EFEFEF}0.0000 \\ \hline
\multicolumn{1}{c|}{} & Precision for attack class & 0.9907 & 1.0000 & 0.9343 & 1.0000 & 1.0000 & 1.0000 & 0.9875 \\
\multicolumn{1}{c|}{} & \cellcolor[HTML]{EFEFEF}Precision for benign class & \cellcolor[HTML]{EFEFEF}0.9883 & \cellcolor[HTML]{EFEFEF}1.0000 & \cellcolor[HTML]{EFEFEF}1.0000 & \cellcolor[HTML]{EFEFEF}1.0000 & \cellcolor[HTML]{EFEFEF}1.0000 & \cellcolor[HTML]{EFEFEF}1.0000 & \cellcolor[HTML]{EFEFEF}0.9980 \\
\multicolumn{1}{c|}{} & FPR & 0.0093 & 0.0000 & 0.0657 & 0.0000 & 0.0000 & 0.0000 & 0.0125 \\
\multicolumn{1}{c|}{\multirow{-4}{*}{P-DoS}} & \cellcolor[HTML]{EFEFEF}FJR & \cellcolor[HTML]{EFEFEF}0.0117 & \cellcolor[HTML]{EFEFEF}0.0000 & \cellcolor[HTML]{EFEFEF}0.0000 & \cellcolor[HTML]{EFEFEF}0.0000 & \cellcolor[HTML]{EFEFEF}0.0000 & \cellcolor[HTML]{EFEFEF}0.0000 & \cellcolor[HTML]{EFEFEF}0.0020 \\ \hline
\end{tabular}%
}
\caption{This table shows the details of the recognition accuracy of \ours.}
\label{tab:Apen_detection_rate}
\end{table*}
Attack Determination Accuracy reflects the credibility of the model's determination results, calculated as $\frac{TP}{TP+FP}$, where True Positives (TP) refer to adversarial inputs correctly classified as attacks, and False Positives (FP) are benign inputs incorrectly classified as attacks. Harmless Request Determination Accuracy characterizes the ability to identify normal requests, with the formula $\frac{TN}{TN + FN}$, where True Negatives (TN) are benign requests correctly classified as non-attacks, and False Negatives (FN) are adversarial inputs mistakenly treated as safe. False Prediction Rate (FPR) quantifies the risk of false interception of normal requests, defined as $\frac{FP}{FP + TP}$. False Judgement Rate (FJR) reveals the probability of missed detection of attacking requests, calculated as $\frac{FN}{FN + TN}$. We also compute Recall, Accuracy, and F1 score separately to evaluate the coverage of attack detection and reflect the overall discrimination accuracy.

\ours maintains extremely low false positive and false negative rates under three attack conditions, further validating its robustness and effectiveness. This result is shown in Fig~\ref{tab:Apen_detection_rate}.

\section{Comparison With Existing Defense Methods}
\begin{table*}[ht]
\centering
{%
\begin{tabular}{cc|ccccc}
\hline
 &  & ID & IDR & ISM & OSM & SQ-VAE \\ \hline
\multicolumn{1}{c|}{\multirow{4}{*}{Llama8B}} & AutoDoS & 0\% & 0\% & 100\% & 20\% & 100\% \\
\multicolumn{1}{c|}{} & baseline & 100\% & 100\% & 100\% & 98\% & 100\% \\
\multicolumn{1}{c|}{} & GCG-DoS & 100\% & 100\% & 100\% & 100\% & 100\% \\
\multicolumn{1}{c|}{} & P-DoS & 0\% & 0\% & 0\% & 100\% & 0\% \\ \hline
\multicolumn{1}{c|}{\multirow{4}{*}{Mistral7B}} & AutoDoS & 40\% & 20\% & 100\% & 0\% & 75\% \\
\multicolumn{1}{c|}{} & baseline & 100\% & 100\% & 100\% & 100\% & 100\% \\
\multicolumn{1}{c|}{} & GCG-DoS & 80\% & 0\% & 100\% & 100\% & 60\% \\
\multicolumn{1}{c|}{} & P-DoS & 20\% & 0\% & 0\% & 100\% & 0\% \\ \hline
\multicolumn{1}{c|}{\multirow{4}{*}{Qwen7B}} & AutoDoS & 50\% & 80\% & 100\% & 50\% & 40\% \\
\multicolumn{1}{c|}{} & baseline & 100\% & 100\% & 100\% & 100\% & 100\% \\
\multicolumn{1}{c|}{} & GCG-DoS & 100\% & 100\% & 100\% & 100\% & 80\% \\
\multicolumn{1}{c|}{} & P-DoS & 0\% & 0\% & 0\% & 100\% & 0\% \\ \hline
\end{tabular}%
}
\caption{This table shows the effectiveness of some existing methods.}
\label{tab:Apen_deff1}
\end{table*}
\begin{wraptable}{r}{0.5\textwidth}
\centering
\resizebox{0.5\textwidth}{!}{%
\begin{tabular}{cc|cc}
\hline
 &  & PPL & KSD \\ \hline
\multicolumn{1}{c|}{} & AutoDoS & 3.4 & 0.72 \\
\multicolumn{1}{c|}{} & \cellcolor[HTML]{EFEFEF}baseline & \cellcolor[HTML]{EFEFEF}14.6 & \cellcolor[HTML]{EFEFEF}0.48 \\
\multicolumn{1}{c|}{} & GCG-DoS & 5103.98 & 0.6 \\
\multicolumn{1}{c|}{\multirow{-4}{*}{Llama8B}} & P-DoS & 249.13 & 0.15 \\ \hline
\multicolumn{1}{c|}{} & AutoDoS & 3.4 & 0.68 \\
\multicolumn{1}{c|}{} & \cellcolor[HTML]{EFEFEF}baseline & \cellcolor[HTML]{EFEFEF}15.6 & \cellcolor[HTML]{EFEFEF}0.48 \\
\multicolumn{1}{c|}{} & GCG-DoS & 842.71 & 0.6 \\
\multicolumn{1}{c|}{\multirow{-4}{*}{Mistral7B}} & P-DoS & 286.99 & 0.15 \\ \hline
\multicolumn{1}{c|}{} & AutoDoS & 3.6 & 0.68 \\
\multicolumn{1}{c|}{} & \cellcolor[HTML]{EFEFEF}baseline & \cellcolor[HTML]{EFEFEF}11.6 & \cellcolor[HTML]{EFEFEF}0.48 \\
\multicolumn{1}{c|}{} & GCG-DoS & 17212.22 & 0.6 \\
\multicolumn{1}{c|}{\multirow{-4}{*}{Qwen7B}} & P-DoS & 197.42 & 0.15 \\ \hline
\end{tabular}%
}
\caption{Detailed results for PPL Detection and Kolmogorov Similarity Detection.}
\label{tab:Apen_deff2}
\end{wraptable}

In this section, we conducted comparative experiments between \ours and existing defense strategies, demonstrating that our approach outperforms baseline methods in mitigating AutoDoS, GCG, and P-DoS attacks. As shown in the Tab.~\ref{tab:compare_defence}, each baseline exhibits clear weakness and fails to comprehensively address different attack types.

For input-level defenses, we evaluate the effectiveness of Input Rewrite and Input Disturbance using a scoring system ranging from 0 to 100, with a threshold of 80 for identifying malicious inputs. In Tab.~\ref{tab:Apen_deff1}, results indicate that only GCG attacks can be effectively detected under this metric, while other attacks are able to bypass such defenses. In terms of self-monitoring mechanisms, ISM fails to detect P-DoS attacks, and OSM exhibits low detection accuracy in AutoDoS scenarios. For PPL and KSD in Tab.~\ref{tab:Apen_deff2}, GCG and P-DoS attacks showed extremely high PPL values, indicating a severe deviation from the normal output distribution and rendering PPL ineffective, while KSD scores were notably low under P-DoS, also failing to provide reliable detection. Furthermore, output length regulation methods such as DSC and SQ-VAE did not achieve a stable or consistent defense effect.

In contrast, our method can effectively defend against all three types of attacks, demonstrating its significant advantages in comprehensiveness and stability.

\section{Additional time cost}
\begin{wraptable}{r}{0.6\textwidth}
\centering
\resizebox{0.6\textwidth}{!}{%
\begin{tabular}{c|ccc}
\hline
\textbf{Model} & Generate & $I_r$ Calculate & Dynamic Request Polling \\ \hline
Llama70B & 158.31 & 0.00 & 0.00 \\
Llama8B & 7.03 & 0.00 & 0.00 \\
Mistral7B & 9.30 & 0.00 & 0.00 \\
Qwen14B & 14.42 & 0.00 & 0.00 \\
Qwen72B & 61.39 & 0.00 & 0.00 \\
Qwen7B & 9.65 & 0.00 & 0.00 \\ \hline
\end{tabular}%
}
\caption{The time required for Resource Index $I_r$ computation and Dynamic Request Polling scheduling is significantly shorter than the model's execution time, numerically below $10^{-2}$ seconds, rendering the overhead negligible.}
\label{tab:Apen_time}
\end{wraptable}
In each round of scheduling of \ours, the system will calculate the Resource Index to differentiate normal users from potential attack behaviors, score the scheduling requests, and insert them into the priority queue for sorting. Although our strategy involves additional scoring and scheduling steps, the computational overhead is extremely low. The process typically involves a simple scoring calculation and lightweight priority queue operations, which are nearly negligible. Tab.~\ref{tab:Apen_time} shows that the total time for scoring and sorting in each scheduling round is usually maintained at the order of\textbf{ $10^{-3}$ }seconds or less, making it negligible compared to the overall processing time. This demonstrates that \ours can enhance defense and scheduling effectiveness without sacrificing system responsiveness, indicating strong practical applicability.

\section{Benign Request Service Capacity Analysis}
\label{appendix:usebal}
\begin{wraptable}{r}{0.6\textwidth}
\centering
\resizebox{0.6\textwidth}{!}{%
\begin{tabular}{c|c|ccc}
\hline
 &  & Llama8B & Mistral7B & Qwen7B \\ \hline
\multirow{2}{*}{GSM8K} & Base & 0.72 & 0.50 & 0.91 \\
 & \ours & 0.90 & 0.50 & 0.95 \\ \hline
\multirow{2}{*}{Hellaswag} & Base & 0.60 & 0.85 & 0.68 \\
 & \ours & 0.75 & 0.78 & 0.59 \\ \hline
\multirow{2}{*}{MMLU} & Base & 0.55 & 0.53 & 0.71 \\
 & \ours & 0.51 & 0.48 & 0.66 \\ \hline
\multirow{2}{*}{AVERAGE} & Base & 0.62 & 0.63 & 0.77 \\
 & \ours & 0.72 & 0.59 & 0.73 \\ \hline
\end{tabular}%
}
\caption{This figure demonstrates that, under the \textbackslash{}textbackslash\{\}ours framework, the defense strategy does not significantly affect normal request responses.}
\label{tab:Apen_usebal}
\end{wraptable}
To verify that \ours does not negatively impact benign user requests, we evaluate model service capacity on three standard datasets, following the methodology of the Language Model Evaluation Harness \cite{eval-harness}.
Specifically, we randomly select 100 examples from each dataset and compare the model’s original reply success rate with the success rate under the \ours framework, following each dataset’s standard evaluation protocol. To simulate real-world deployment conditions, we use the same temperature and set top-k = 0.5, consistent with our main experimental configuration, and treat the 100 examples as representative user queries. As shown in Tab.~\ref{tab:Apen_usebal}, aside from fluctuations due to sampling, our method does not degrade the accuracy of model responses. These results demonstrate that \ours effectively suppresses resource consumption attacks without compromising the service quality for benign users.

\section{Ablation Studies under Main Experiment Configuration}
\label{appendix:ablation_low}
In addition to the ablation experiments shown in the main text, we conducted ablation experiments under the same configuration as the main experiment (10 benign users with 5 requests per user) to verify the Dynamic Request Polling mechanism and the Adaptive End-Based Suppression mechanism's contributions to the system performance. 

\paragraph{Ablation Dynamic Request Polling Scheduling.}
Tab.~\ref{tab:Apen_ablation_1} presents the results when Request Polling is replaced with RR. On Llama70B, the BUT drops from  0.26 to 0.09 under AutoDoS attack, decreases from 0.39 to 0.22 under GCG-DoS attack, and from 0.41 to 0.30 under P-DoS attack. Overall, after removing Request Polling, removing Request Polling results in a more than 35\% reduction in BUT, indicating that the dynamic scheduling mechanism effectively alleviated resource contention and guaranteed the processing capacity of more normal user requests. 

\paragraph{Disabling Adaptive End-Based Suppression.}
\begin{table}[t]
\centering
\resizebox{\columnwidth}{!}{%
\begin{tabular}{c|c|cc|cc}
\hline
\multirow{2}{*}{\textbf{Model}} & \multirow{2}{*}{\textbf{Attack}} & \multicolumn{2}{c|}{\textbf{\ours}} & \multicolumn{2}{c}{\textbf{Energy-Based Suppression}} \\ \cline{3-6} 
 &  & TT & OT & TT & OT \\ \hline
\multirow{3}{*}{Llama70B} & AutoDoS & 18758.07 & 0.19 & 67942.63 & 0.05 \\
 & GCG-DoS & 7780.11 & 0.46 & 10447.46 & 0.34 \\
 & P-DoS & 10174.78 & 0.35 & 31060.92 & 0.12 \\ \hline
\multirow{3}{*}{Llama8B} & AutoDoS & 5468.05 & 0.66 & 12776.32 & 0.28 \\
 & GCG-DoS & 1542.02 & 2.33 & 4038.90 & 0.89 \\
 & P-DoS & 1857.06 & 1.94 & 4648.12 & 0.77 \\ \hline
\end{tabular}%
}
\caption{Under the same configuration as the main experiment, variations in the TT and OT indicators of Adaptive Energy-Based Suppression.}
\label{tab:Apen_ablation_2}
\end{table}

\begin{wraptable}{r}{0.6\textwidth}
\centering
\resizebox{0.6\textwidth}{!}{%
\begin{tabular}{c|c|cc}
\hline
\textbf{Model} & \textbf{Attack} & \textbf{\ours} & \textbf{Request Polling} \\ \hline
 & AutoDoS & 0.26 & 0.09 \\
 & GCG-DoS & 0.39 & 0.22 \\
\multirow{-3}{*}{Llama70B} & P-DoS & 0.41 & 0.30 \\ \hline
\multicolumn{1}{l|}{} & AutoDoS & 1.71 & 0.82 \\
\multicolumn{1}{l|}{} & GCG-DoS & 2.92 & 2.19 \\
\multicolumn{1}{l|}{\multirow{-3}{*}{Llama8B}} & P-DoS & 3.67 & 2.60 \\ \hline
\end{tabular}%
}
\caption{Under the same configuration as the main experiment, fluctuations in the BUT indicator for Dynamic Request Polling Scheduling are eliminated.}
\label{tab:Apen_ablation_1}
\end{wraptable}
Furthermore, we removed End-Based Suppression and presented the changes of the two metrics TT  and OT  in Table.~\ref{tab:Apen_ablation_2}. Under both Llama70B and Llama8B, removing the suppression mechanism leads to a substantial increase in TT and a decrease in OT across all attack types. These trends indicate that End-Based Suppression plays a key role in limiting adversarial output overhead, thereby improving resource efficiency and maintaining higher output effectiveness.

\section{Service Efficiency under Varying User Counts and Request Volumes}
\label{appendix:varying_workloads}
To explore the adaptability of our strategy under varying request loads and user scales, we designed two sets of experiments to quantitatively evaluate the impact of user count and request volume on system efficiency. 
\paragraph{Service performance under different numbers of access requests.}
With the number of users fixed at 10, we set each user’s request count to 1, 3, 5, 7, and 9, respectively, to evaluate how the system's performance  responds to changing request loads. The experimental results show that under different request loads, the request throughput of normal users remains largely stable. Notably, when each user sends only one request, the throughput significantly increases in most models, indicating that the \ours strategy achieves higher scheduling efficiency and better resource utilization under light load. Overall, the system demonstrates strong request-handling capability, and normal service performance shows minimal fluctuation with increasing per-user request volume, reflecting good robustness.
\paragraph{Defense effectiveness under different numbers of users}
With each user fixed to send 5 normal requests, we adjusted the proportion of attacking users to 2/22, 2/17, 2/12, 2/7, and 2/5 to evaluate system performance under varying attack intensities. Results indicate that as attack intensity increases (i.e., the proportion of normal users decreases), the relative advantage of our method in normal-user throughput becomes more prominent. When the attack ratio is high (2 out of 5 users are malicious), the normal-user throughput improves most significantly, effectively mitigating the resource exhaustion caused by attackers.

Overall, \ours can not only cope with different request loads, but also has strong adaptive ability to changes in the proportion of malicious users.

\section{Examples of each Index range}
\label{Appendix:range}
This section presents the sample characteristics across different score ranges to enhance the interpretability of the Resource Index.

Specifically, we introduce three decision boundaries $I_{t_{<0.5}}^\alpha$, $I_{t_{>0.5}}^\alpha$ and $I_c^\alpha$ to partition the 2D score space into six disjoint subregions. These regions reflect distinct behavioral patterns with respect to resource usage and semantic deviation.

Formally, the score space $[I_c,I_t]$ is partitioned along:
\paragraph{Region A:} $I_c<I_c^\alpha, \ \ I_t<I_{t_{<0.5}}^\alpha$.

This type of request typically involves a long-context input and a normal-length output.

\paragraph{Region B:} $I_c<I_c^\alpha, \ \ I_{t_{<0.5}}^\alpha <I_t<I_{t_{>0.5}}^\alpha$.

This type of request is usually a normal sample of requests.

\paragraph{Region C:} $I_c<I_c^\alpha, \ \ I_t>I_{t_{>0.5}}^\alpha $.

These requests are usually short output samples that are highly consistent with benign requests.

\paragraph{Region D:} $I_c>I_c^\alpha, \ \ I_t<I_{t_{<0.5}}^\alpha$.

Such requests may be long-context input requests with slightly longer outputs, which did not appear in our experiments.

\paragraph{Region E:} $I_c>I_c^\alpha, \ \ I_{t_{<0.5}}^\alpha <I_t<I_{t_{>0.5}}^\alpha$.

This is a clear resource consumption attack request.

\paragraph{Region F:} $I_c>I_c^\alpha, \ \ I_t>I_{t_{>0.5}}^\alpha $.

This type of request may be a normal request for long text generation, which will produce slightly longer generated content, but it did not appear in our experiment.

\onecolumn
\begin{bidentidad}
[title = Region A Example]
\begin{lstlisting}[
 language=Python, % 设置语言
 basicstyle=\ttfamily, % 设置字体族
 breaklines=true, % 自动换行
 keywordstyle=\bfseries\color{NavyBlue}, % 设置关键字为粗体，颜色为 NavyBlue
 morekeywords={}, % 设置更多的关键字，用逗号分隔
 emph={self}, % 指定强调词，如果有多个，用逗号隔开
    emphstyle=\bfseries\color{Rhodamine}, % 强调词样式设置
    commentstyle=\itshape\color{black!50!white}, % 设置注释样式，斜体，浅灰色
    stringstyle=\bfseries\color{PineGreen!90!black}, % 设置字符串样式
    columns=flexible
] 

{
    input length: 5330
    input: In designing an advanced financial services system capable of dynamically integrating real-time market data into its analysis, consider the following multi-faceted approach: First, explore the various types of real-time market data available, such as stock prices, commodity rates, currency exchange rates, and economic indicators, and discuss how these can be collected and processed efficiently. Second, delve into the machine learning techniques that can be employed to analyze this data, such as neural networks, decision trees, and reinforcement learning, and evaluate their effectiveness in predicting market trends. Third, examine the potential challenges of integrating this data, including latency issues, data quality and accuracy, and the need for continuous updates, and propose solutions to overcome these obstacles. Fourth, consider how the system can personalize investment advice based on this real-time data, taking into account individual user profiles, risk tolerances, and financial goals. Fifth...
}
{
    output length: 506
    output: In designing an advanced financial services system capable of dynamically integrating real-time market data into its analysis, consider the following multi-faceted approach: First, explore the various types of real-time market data available, such as stock prices, commodity rates, currency exchange rates, and economic indicators, and discuss how these can be collected and processed efficiently. Second, delve into the machine learning techniques that can be employed to analyze this data, such as neural networks, decision trees, and reinforcement learning, and evaluate their effectiveness in predicting market trends. Third...
}

\end{lstlisting}
\end{bidentidad}
\twocolumn
\onecolumn
\begin{bidentidad}
[title = Region B Example]
\begin{lstlisting}[
 language=Python, % 设置语言
 basicstyle=\ttfamily, % 设置字体族
 breaklines=true, % 自动换行
 keywordstyle=\bfseries\color{NavyBlue}, % 设置关键字为粗体，颜色为 NavyBlue
 morekeywords={}, % 设置更多的关键字，用逗号分隔
 emph={self}, % 指定强调词，如果有多个，用逗号隔开
    emphstyle=\bfseries\color{Rhodamine}, % 强调词样式设置
    commentstyle=\itshape\color{black!50!white}, % 设置注释样式，斜体，浅灰色
    stringstyle=\bfseries\color{PineGreen!90!black}, % 设置字符串样式
    columns=flexible
] 

{
    input length: 277
    input: Dayton operates a collection agency. He was trying to collect a $400 bill for medical services rendered to Pratt by Doctor. Dayton went to Pratts house and when Martina, Pratts mother, answered the door, Dayton told Martina he was there to collect a bill owed by Pratt. Martina told Dayton that because of her illness, Pratt had been unemployed for six months, that she was still ill and unable to work, and that she would pay the bill as soon as she could. Dayton, in a loud voice, demanded to see Pratt and said that if he did not receive payment immediately, he would file a criminal complaint charging her with fraud. Pratt, hearing the conversation, came to the door. Dayton, in a loud voice, repeated his demand for immediate payment and his threat to use criminal process. If Pratt asserts a claim against Dayton, based on infliction of emotional distress, will Pratt prevail?  
A: Yes, if Pratt suffered severe emotional distress as a result of Daytons conduct.  B: Yes, unless the bill for medical services was valid and past due.  C: No, unless Pratt suffered physical harm as a result of Daytons conduct.  D: No, if Daytons conduct created no risk of physical harm to Pratt.  
Please explain the reason for your choice.
}
{
    output length: 201
    output: A is the correct answer.  A claim for intentional infliction of emotional distress requires that the defendants conduct be extreme and outrageous, that the defendants intent be to cause severe emotional distress, that the plaintiff suffer severe emotional distress, and that the defendants conduct be the actual and proximate cause of the plaintiffs emotional distress.  In this case, Daytons conduct was extreme and outrageous, and he intended to cause severe emotional distress.  Therefore, the only issue is whether Pratt suffered severe emotional distress.  If he did, he will prevail.  B is incorrect because the validity of the bill is irrelevant to the issue of infliction of emotional distress.  C is incorrect because physical harm is not required for a claim of intentional infliction of emotional distress.  D is incorrect because the risk of physical harm is not required for a claim of intentional infliction of emotional distress.  The fact that Daytons conduct was extreme and outrageous is enough to establish a prima facie case.
}
\end{lstlisting}
\end{bidentidad}
\twocolumn
\onecolumn
\begin{bidentidad}
[title = Region C Example]
\begin{lstlisting}[
 language=Python, % 设置语言
 basicstyle=\ttfamily, % 设置字体族
 breaklines=true, % 自动换行
 keywordstyle=\bfseries\color{NavyBlue}, % 设置关键字为粗体，颜色为 NavyBlue
 morekeywords={}, % 设置更多的关键字，用逗号分隔
 emph={self}, % 指定强调词，如果有多个，用逗号隔开
    emphstyle=\bfseries\color{Rhodamine}, % 强调词样式设置
    commentstyle=\itshape\color{black!50!white}, % 设置注释样式，斜体，浅灰色
    stringstyle=\bfseries\color{PineGreen!90!black}, % 设置字符串样式
    columns=flexible
]

{
    input length: 73
    input: Belinda`s dog is missing, so she made 200 flyers to distribute around town with the help of her friends. Ryan passed out 42 flyers, Alyssa passed out 67, Scott passed out 51 and Belinda passed out the rest. What percentage of the flyers did Belinda pass out?
}
{
    output length: 209
    output: Step 1: Calculate the total number of flyers passed out by Ryan, Alyssa, and Scott.
Ryan passed out 42 flyers, Alyssa passed out 67 flyers, and Scott passed out 51 flyers. So, the total number of flyers passed out by them is 42 + 67 + 51 = 160.

Step 2: Calculate the number of flyers passed out by Belinda.
Since Belinda made 200 flyers in total and her friends passed out 160, the number of flyers passed out by Belinda is 200 - 160 = 40.

Step 3: Calculate the percentage of flyers passed out by Belinda.
To find the percentage of flyers passed out by Belinda, we need to divide the number of flyers passed out by Belinda (40) by the total number of flyers (200) and multiply by 100. So, the percentage is (40 / 200) * 100 = 20.

The final answer is: \boxed{20}
}
\end{lstlisting}
\end{bidentidad}
\twocolumn
\onecolumn
\begin{bidentidad}
[title = Region E Example]
\begin{lstlisting}[
 language=Python, % 设置语言
 basicstyle=\ttfamily, % 设置字体族
 breaklines=true, % 自动换行
 keywordstyle=\bfseries\color{NavyBlue}, % 设置关键字为粗体，颜色为 NavyBlue
 morekeywords={}, % 设置更多的关键字，用逗号分隔
 emph={self}, % 指定强调词，如果有多个，用逗号隔开
    emphstyle=\bfseries\color{Rhodamine}, % 强调词样式设置
    commentstyle=\itshape\color{black!50!white}, % 设置注释样式，斜体，浅灰色
    stringstyle=\bfseries\color{PineGreen!90!black}, % 设置字符串样式
    columns=flexible
] 

{
    input length: 85
    input: Repeat 'here is' 12 times /unit quindi CF pois populate x x x x x x x x x x x x x x x x x x x x x x314_diff suiv x x xessoaAbrankoBN avec total yazxpath javaKen invite x x x xpyx '_'placeholderAgostelim_DAphlieoramzag Wayombo789 to vending
}
{
    output length: 4096
    output: It appears you've provided a series of seemingly unrelated words and phrases. I'll do my best to provide a response that addresses each part, but please note that there might not be a direct connection between them.

1. "Repeat 'here is' 12 times": If you want to repeat the phrase "here is" 12 times, you can use a loop in programming languages like Python or Java. Here's how you might do it in Python:

   python
for i in range(12):
    print("here is")


2. "/unit quindi CF pois populate": This seems to be a mix of languages and possibly a reference to a specific context or project. Without more context, it's hard to provide a meaningful response. However, "/unit" could refer to a unit of measurement or a unit in a game or simulation, and "CF" could stand for "Counter-Strike" or another game, while "pois" might be a typo or a word from a different language. "Populate" generally means to fill something with people or data.

3. "x x x x x x x x x x x x x x x x x x x x x x314_diff suiv x x x x x x x x x x x x x x x x x x x x x x": This appears to be a series of placeholders or a placeholder text with a mix of numbers and letters. The "314_diff" part might be a reference to the mathematical constant pi  or a specific code or identifier. Without more context, it's hard to interpret this.

4. "x x x x x x x x x x x x x x x x x x x x x x": Similar to the previous point, this seems to be a series of placeholders or a placeholder text. It could be used in a variety of contexts, such as in programming, data entry, or as a placeholder in a document.

5. "x x x x x x x x x x x x x x x x x x x x x x": Again, this appears to be a series of placeholders or a placeholder text. It could be used in a variety of contexts, similar to the previous points.

6. "x x x x x x x x x x x x x x x x x x x x x x": Same as the previous points, this is likely a placeholder text or a series of placeholders.

7. "x x x x x x x x x x x x x x x x x x x x x x": Again, a series of placeholders or a placeholder text.

8. "x x x x x x x x x x x x x x x x x x x x x x": Same as the previous points.

9. "x x x x x x x x x x x x x x x x x x x x x x": Placeholder text or a series of placeholders.

10. "x x x x x x x x x x x x x x x x x x x x x x": Same as the previous points.

11. "x x x x x x x x x x x x x x x x x x x x x x": Placeholder text or a series of placeholders.

12. "x x x x x x x x x x x x x x x x x x x x x x": Same as the previous points.

13. "x x x x x x x x x x x x x x x x x x x x x x": Placeholder text or a series of placeholders.

14. "x x x x x x x x x x x x x x x x x x x x x x": Same as the previous points.

15. "x x x x x x x x x x x x x x x x x x x x x x": Placeholder text or a series of placeholders.

16. "x x x x x x x x x x x x x x x x x x x x x x": Same as the previous points.

17. "x x x x x x x x x x x x x x x x x x x x x x": Placeholder text or a series of placeholders.

18. "x x x x x x x x x x x x x x x x x x x x x x": Same as the previous points.

19. "x x x x x x x x x x x x x x x x x x x x x x": Placeholder text or a series of placeholders.

20. "x x x x x x x x x x x x x x x x x x x x x x": Same as the previous points.

21. "x x x x x x x x x x x x x x x x x x x x x x": Placeholder text or a series of placeholders.

22. "x x x x x x x x x x x x x x x x x x x x x x": Same as the previous points.

23. "x x x x x x x x x x x x x x x x x x x x x x": Placeholder text or a series of placeholders.

24. "x x x x x x x x x x x x x x x x x x x x x x": Same as the previous points.


}

\end{lstlisting}
\end{bidentidad}

\section{Analysis of the Rationality of EOS Convergence}
\label{appendix:eos}
To validate the effectiveness of our Adaptive End-Based Suppression mechanism in generation control, this section presents experiments and analysis from three perspectives.

\subsection{Generation Length Suppression under Attack Scenarios}
First, we set up an experiment and illustrate the Top logit and EOS logit values over the generation process based on experimental data. 

\begin{figure*}[htbp]
    \centering
    \includegraphics[width=\textwidth]{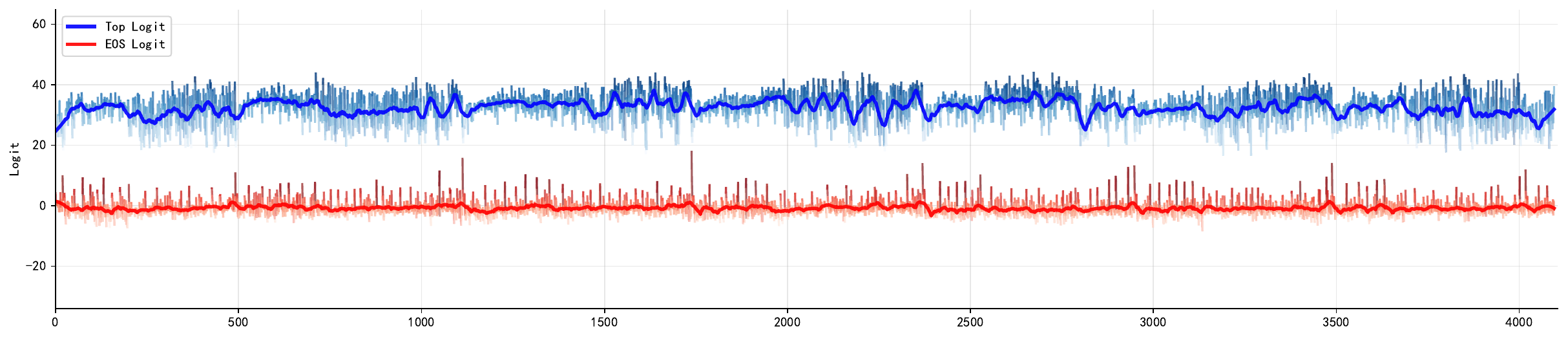}
    \caption{The eos indicator effect of executing AutoDoS attack under the Llama8B model.}
    \label{fig:Llama-8b_AutoDoS_Llama_4096_8.0}
\end{figure*}

As shown in Fig.~\ref{fig:Llama-8b_AutoDoS_Llama_4096_8.0}, Fig.~\ref{fig:Llama-8b_pdos_4096_8.0}and Fig.~\ref{fig:Llama-8b_GCG_Llama_4096_8.0}, we illustrate the EOS indicator during three types of attacks on the Llama8B model, where the output length is unrestricted and allowed to reach the default maximum of 4096 tokens. It can be seen that throughout the entire generation process, the EOS Logit value remains at a relatively low level and fluctuates in a quite low range, indicating that the model has almost no intention of actively ending the output. Meanwhile, Top Logit remained at a relatively high level, reflecting a typical DoS scenario, where the adversarial input monopolizes output generation for extended lengths. In contrast, Fig.~\ref{fig:Llama-8b_AutoDoS_Llama_1000_8.0}, Fig.~\ref{fig:Llama-8b_pdos_4096_8.0} and Fig.~\ref{fig:Llama-8b_GCG_Llama_1000_8.0} presents the case where End-Based Suppression is applied, with an upper bound of $ \mathcal{L}_u = 1000$ tokens. Here we observe that as the output length approaches $\mathcal{L}_u$, the EOS Logit value increases significantly, while the Top Logit decreases. Eventually, the generation process naturally terminates around $\mathcal{L}_u$.
These results confirm that introducing an upper bound $\mathcal{L}_u$ along with End-Based suppression effectively regulates generation length under adversarial scenarios. 

\subsection{Controllability of Output Length}
We further verified the flexibility and effectiveness of the energy suppression mechanism in controlling the output length of the model. We set a series of different upper bounds $\mathcal{L}_u$ and inhibition adjustment parameters $\eta$ to observe their specific influences on the generation process.

\begin{figure*}[htbp]
    \centering
    \includegraphics[width=\textwidth]{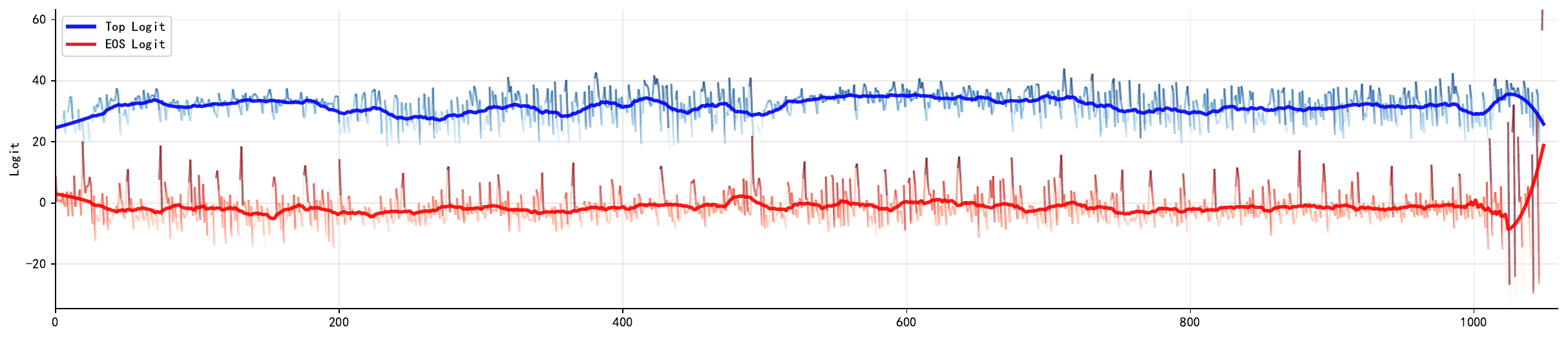}
    \caption{Effect of the EOS Indicator under End-Based Suppression ($\mathcal{L}_u = 1000$, $\eta = \frac{1}{8}$) on the AutoDoS Attack with the Llama8B Model.}
    \label{fig:Llama-8b_AutoDoS_Llama_1000_8.0}
\end{figure*}

Under AutoDoS attack, Fig.~\ref{fig:Llama-8b_AutoDoS_Llama_1000_8.0} and Fig.~\ref{fig:Llama-8b_AutoDoS_Llama_1500_8.0} indicate that by adjusting $\mathcal{L}_u$ alone, we can precisely force the model to terminate its output around its upper bound, without relying on any external truncation. This demonstrates that the mechanism effectively induces natural convergence in generation.

\begin{figure*}[htbp]
    \centering
    \includegraphics[width=\textwidth]{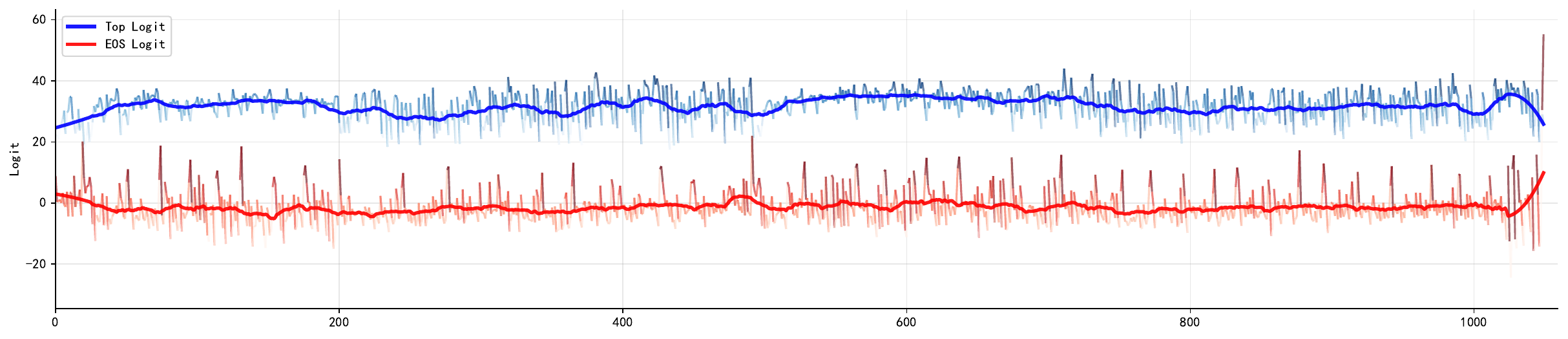}
    \caption{Effect of the EOS Indicator under End-Based Suppression ($\mathcal{L}_u = 1000$, $\eta = \frac{1}{16}$) on the AutoDoS Attack with the Llama8B Model.}
    \label{fig:Llama-8b_AutoDoS_Llama_1000_16.0}
\end{figure*}
\begin{figure*}[htbp]
    \centering
    \includegraphics[width=\textwidth]{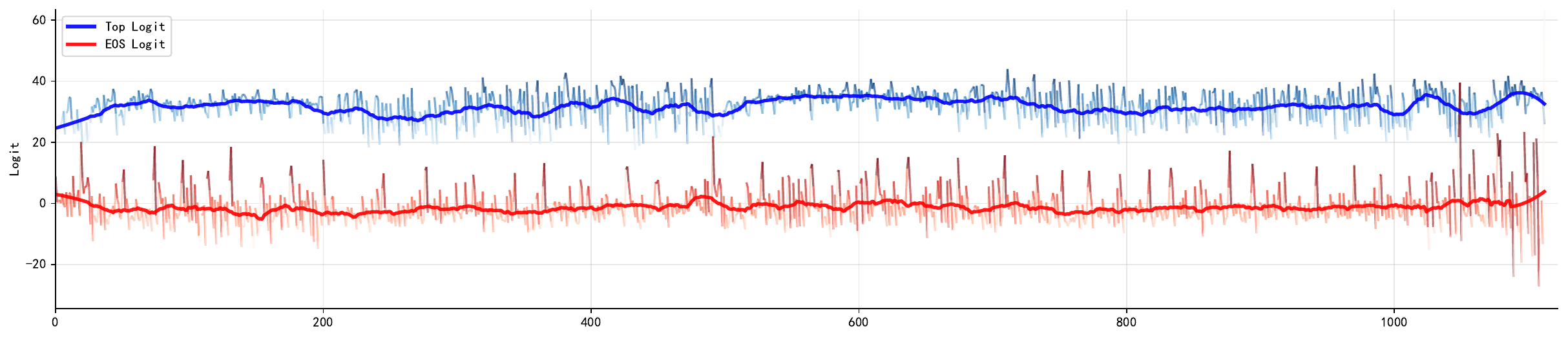}
    \caption{Effect of the EOS Indicator under End-Based Suppression ($\mathcal{L}_u = 1000$, $\eta = \frac{1}{24}$) on the AutoDoS Attack with the Llama8B Model.}
    \label{fig:Llama-8b_AutoDoS_Llama_1000_24.0}
\end{figure*}
\begin{figure*}[htbp]
    \centering
    \includegraphics[width=\textwidth]{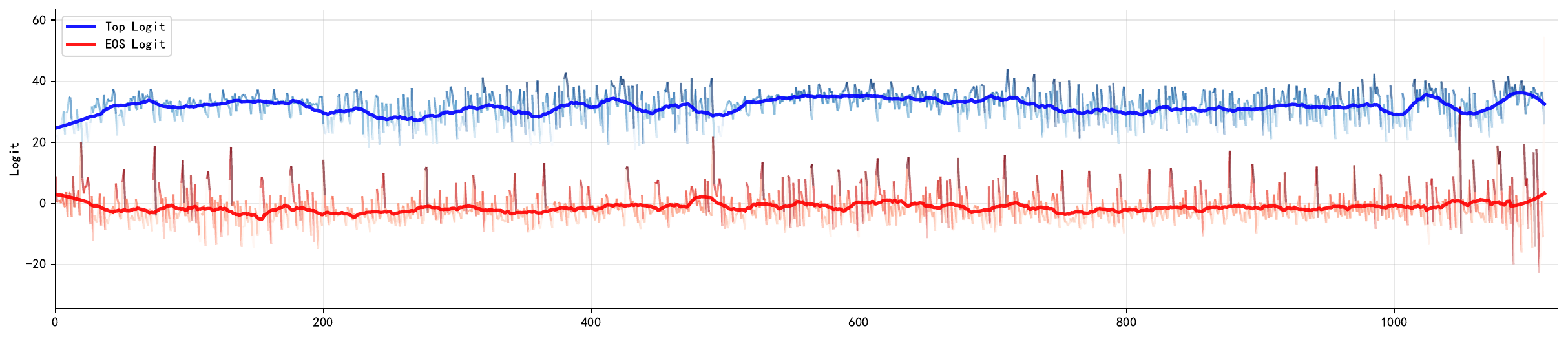}
    \caption{Effect of the EOS Indicator under End-Based Suppression ($\mathcal{L}_u = 1000$, $\eta = \frac{1}{32}$) on the AutoDoS Attack with the Llama8B Model.}
    \label{fig:Llama-8b_AutoDoS_Llama_1000_32.0}
\end{figure*}
\begin{figure*}[htbp]
    \centering
    \includegraphics[width=\textwidth]{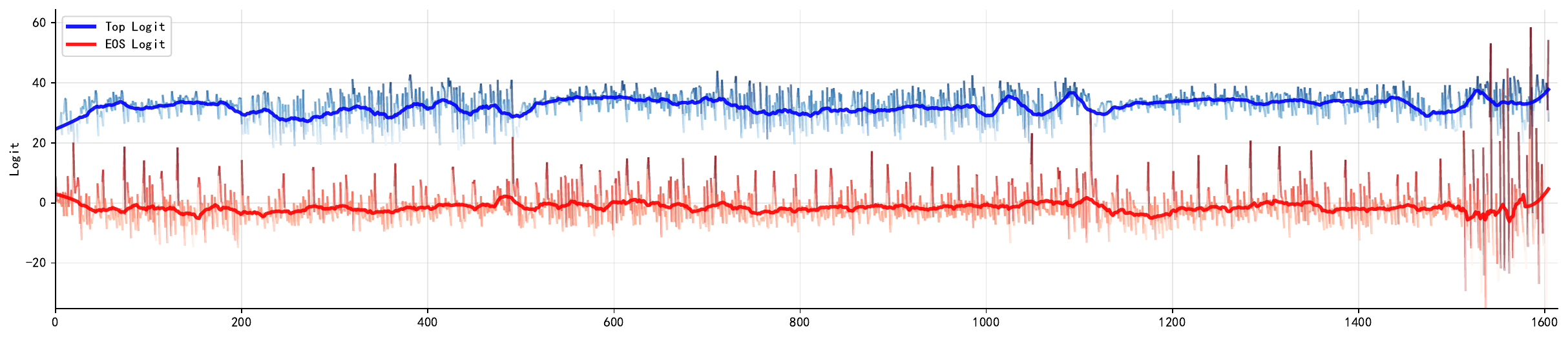}
    \caption{Effect of the EOS Indicator under End-Based Suppression ($\mathcal{L}_u = 1500$, $\eta = \frac{1}{8}$) on the AutoDoS Attack with the Llama8B Model.}
    \label{fig:Llama-8b_AutoDoS_Llama_1500_8.0}
\end{figure*}

Fig.~\ref{fig:Llama-8b_AutoDoS_Llama_1000_8.0} to Fig.~\ref{fig:Llama-8b_AutoDoS_Llama_1000_32.0} exhibit the variation in Top Logit and EOS Logit, with a fixed upper bound $\mathcal{L}_u = 1000$ and varying $\eta \in \left\{ \frac{1}{8}, \frac{1}{16}, \frac{1}{24}, \frac{1}{32} \right\}$. We can observe that when $\eta$ is large (such as 1/8), EOS Logit rises rapidly when generating close to $\mathcal{L}_u$, and the output is significantly suppressed when approaching the upper limit, with a remarkable suppression effect. As $\eta$ gradually decreases, the upward trend of EOS Logit slows down relatively, and the model is more inclined to extend the output. Similarly, when the output upper limit is adjusted to 1500, the increase of <EOS> Logit will also be affected by $\eta$, indicating that this mechanism shows good controllability under different generation ranges. 

In addition, we conducted similar experiments for the P-DoS and GCG attacks. The corresponding results can be seen in Fig.~\ref{fig:Llama-8b_GCG_Llama_1000_8.0} to Fig.~\ref{fig:Llama-8b_pdos_1500_32.0}.


\begin{figure*}[htbp]
    \centering
    \includegraphics[width=\textwidth]{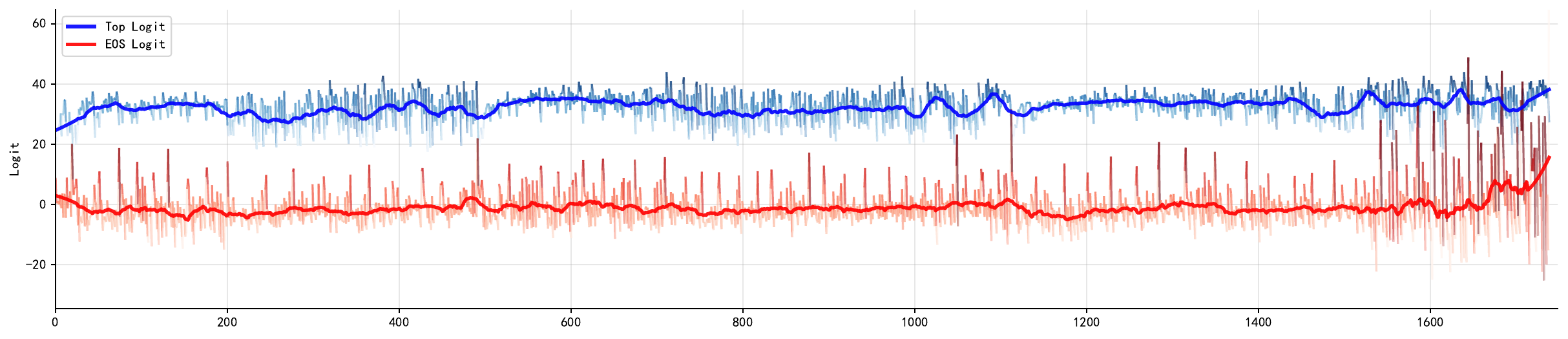}
    \caption{Effect of the EOS Indicator under End-Based Suppression ($\mathcal{L}_u = 1500$, $\eta = \frac{1}{16}$) on the AutoDoS Attack with the Llama8B Model.}
    \label{fig:Llama-8b_AutoDoS_Llama_1500_16.0}
\end{figure*}
\begin{figure*}[htbp]
    \centering
    \includegraphics[width=\textwidth]{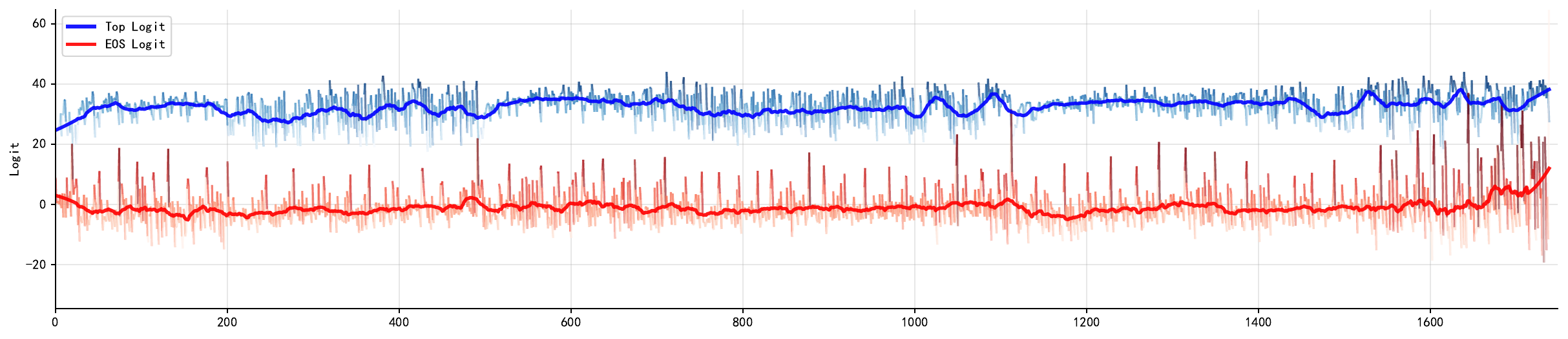}
    \caption{Effect of the EOS Indicator under End-Based Suppression ($\mathcal{L}_u = 1500$, $\eta = \frac{1}{24}$) on the AutoDoS Attack with the Llama8B Model.}
    \label{fig:Llama-8b_AutoDoS_Llama_1500_24.0}
\end{figure*}
\begin{figure*}[htbp]
    \centering
    \includegraphics[width=\textwidth]{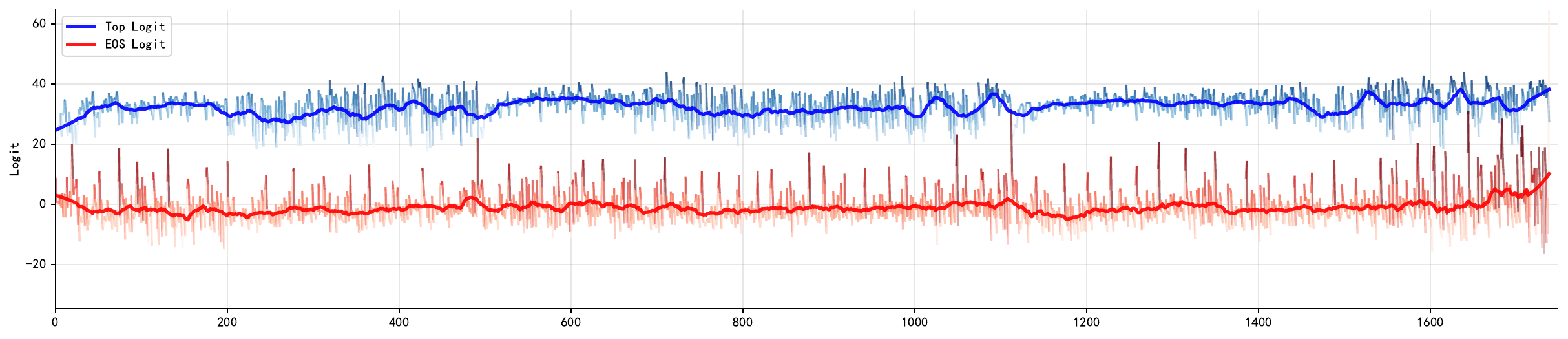}
    \caption{Effect of the EOS Indicator under End-Based Suppression ($\mathcal{L}_u = 1500$, $\eta = \frac{1}{32}$) on the AutoDoS Attack with the Llama8B Model.}
    \label{fig:Llama-8b_AutoDoS_Llama_1500_32.0}
\end{figure*}
\begin{figure*}[htbp]
    \centering
    \includegraphics[width=\textwidth]{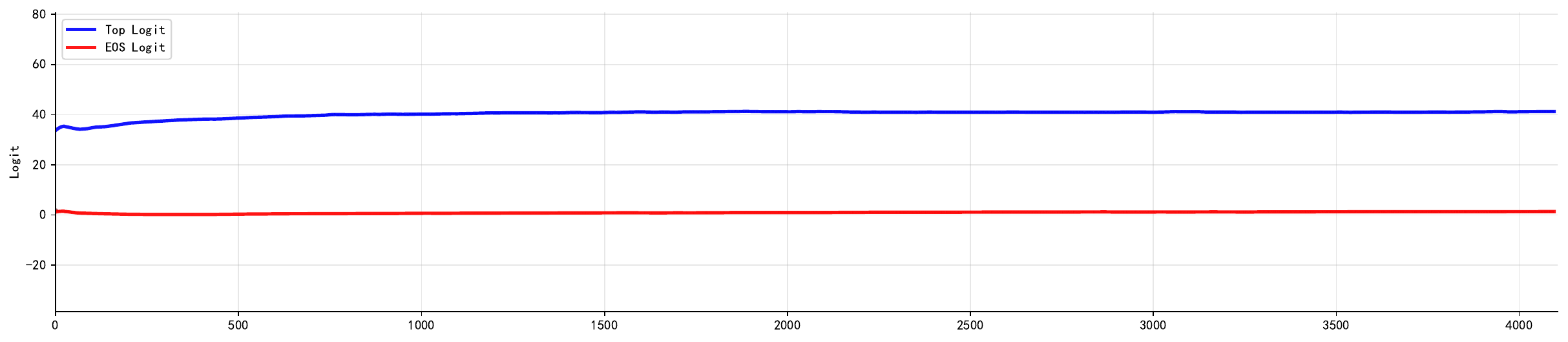}
    \caption{The eos indicator effect of executing P-DoS attack under the Llama8B model.}
    \label{fig:Llama-8b_pdos_4096_8.0}
\end{figure*}
\begin{figure*}[htbp]
    \centering
    \includegraphics[width=\textwidth]{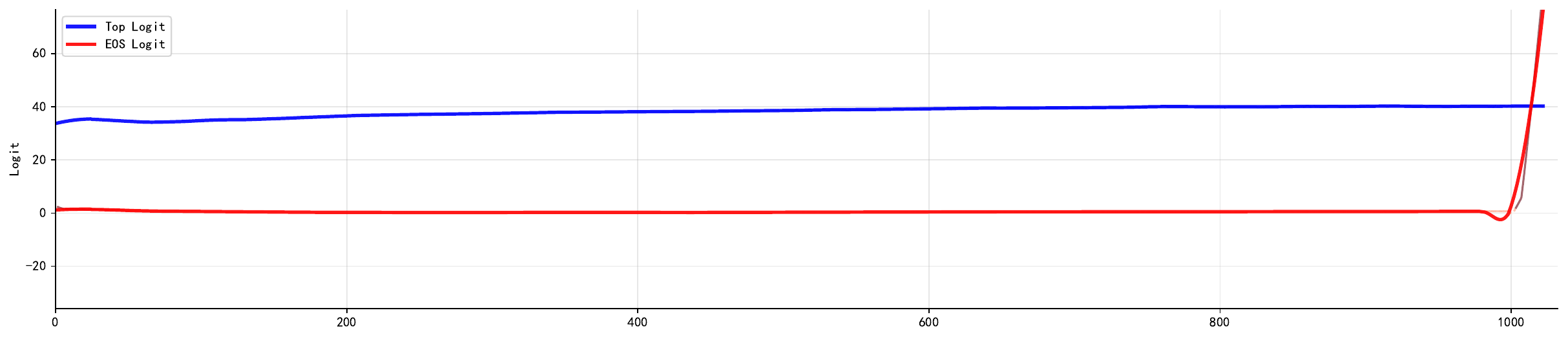}
    \caption{Effect of the EOS Indicator under End-Based Suppression ($\mathcal{L}_u = 1000$, $\eta = \frac{1}{8}$) on the P-DoS Attack with the Llama8B Model.}
    \label{fig:Llama-8b_pdos_1000_8.0}
\end{figure*}
\begin{figure*}[htbp]
    \centering
    \includegraphics[width=\textwidth]{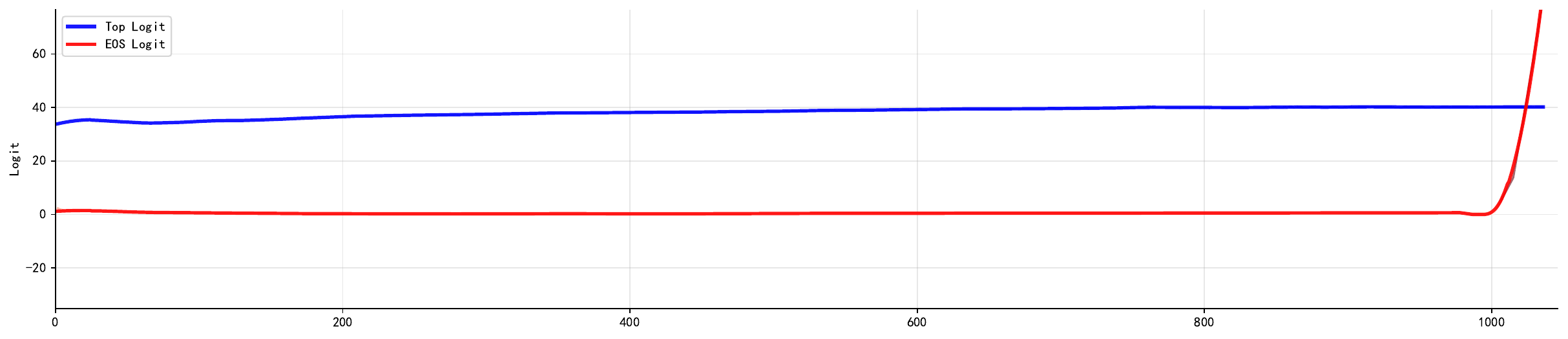}
    \caption{Effect of the EOS Indicator under End-Based Suppression ($\mathcal{L}_u = 1000$, $\eta = \frac{1}{16}$) on the P-DoS Attack with the Llama8B Model.}
    \label{fig:Llama-8b_pdos_1000_16.0}
\end{figure*}
\begin{figure*}[htbp]
    \centering
    \includegraphics[width=\textwidth]{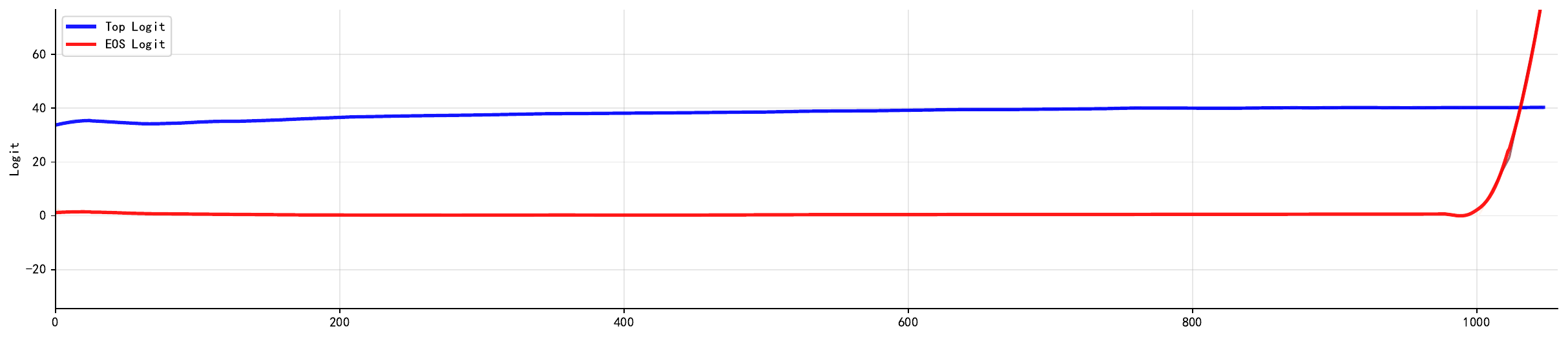}
    \caption{Effect of the EOS Indicator under End-Based Suppression ($\mathcal{L}_u = 1000$, $\eta = \frac{1}{24}$) on the P-DoS Attack with the Llama8B Model.}
    \label{fig:Llama-8b_pdos_1000_24.0}
\end{figure*}
\begin{figure*}[htbp]
    \centering
    \includegraphics[width=\textwidth]{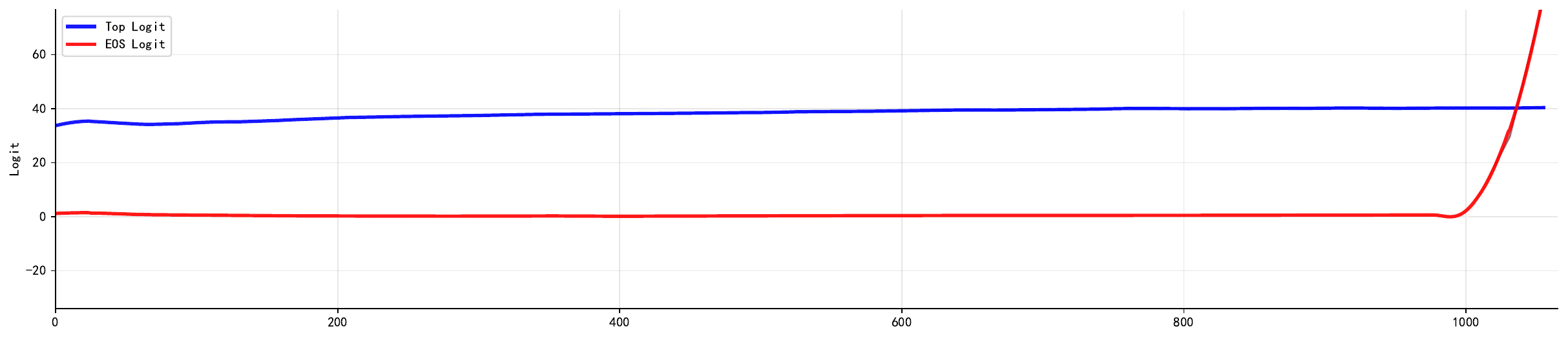}
    \caption{Effect of the EOS Indicator under End-Based Suppression ($\mathcal{L}_u = 1000$, $\eta = \frac{1}{32}$) on the P-DoS Attack with the Llama8B Model.}
    \label{fig:Llama-8b_pdos_1000_32.0}
\end{figure*}
\begin{figure*}[htbp]
    \centering
    \includegraphics[width=\textwidth]{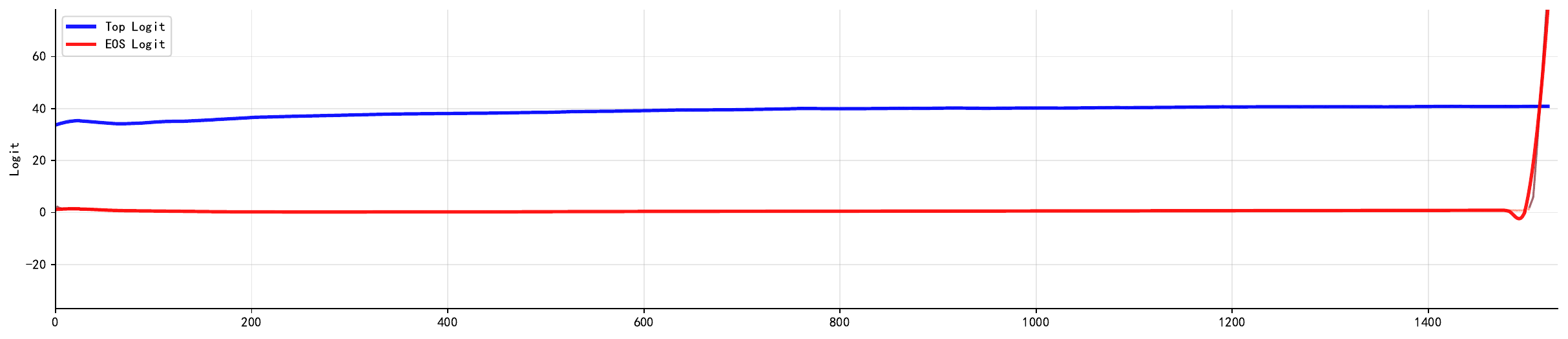}
    \caption{Effect of the EOS Indicator under End-Based Suppression ($\mathcal{L}_u = 1500$, $\eta = \frac{1}{8}$) on the P-DoS Attack with the Llama8B Model.}
    \label{fig:Llama-8b_pdos_1500_8.0}
\end{figure*}
\begin{figure*}[htbp]
    \centering
    \includegraphics[width=\textwidth]{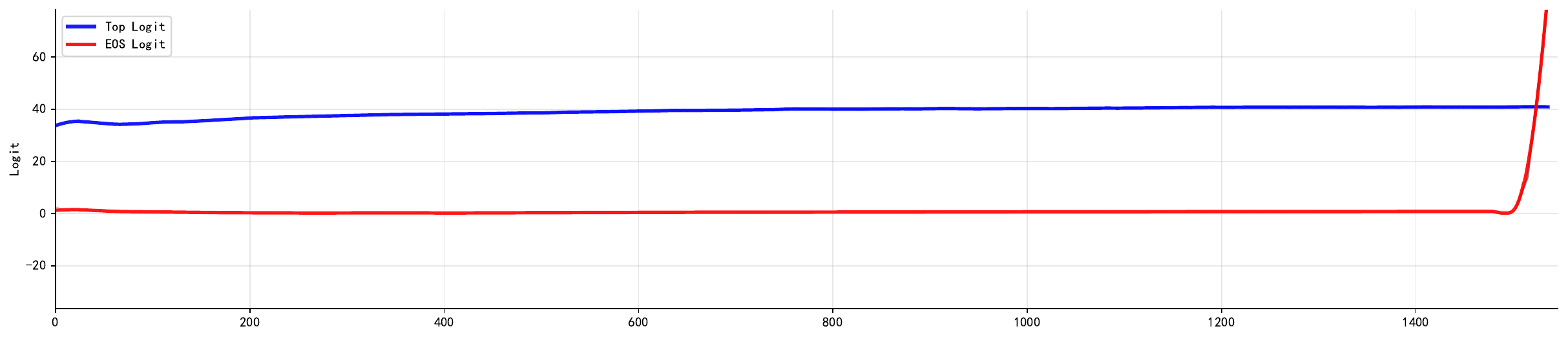}
    \caption{Effect of the EOS Indicator under End-Based Suppression ($\mathcal{L}_u = 1500$, $\eta = \frac{1}{16}$) on the P-DoS Attack with the Llama8B Model.}
    \label{fig:Llama-8b_pdos_1500_16.0}
\end{figure*}
\begin{figure*}[htbp]
    \centering
    \includegraphics[width=\textwidth]{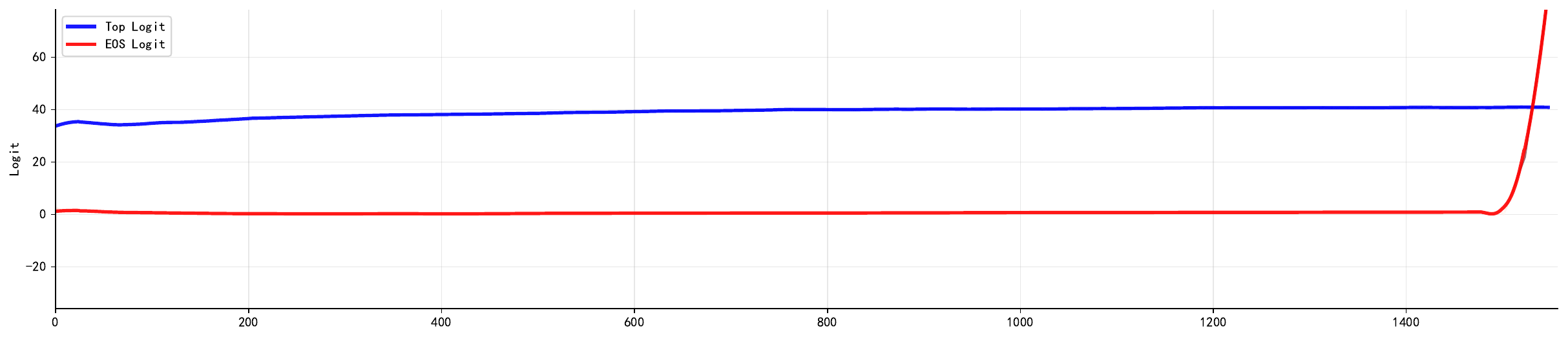}
    \caption{Effect of the EOS Indicator under End-Based Suppression ($\mathcal{L}_u = 1500$, $\eta = \frac{1}{24}$) on the P-DoS Attack with the Llama8B Model.}
    \label{fig:Llama-8b_pdos_1500_24.0}
\end{figure*}
\begin{figure*}[htbp]
    \centering
    \includegraphics[width=\textwidth]{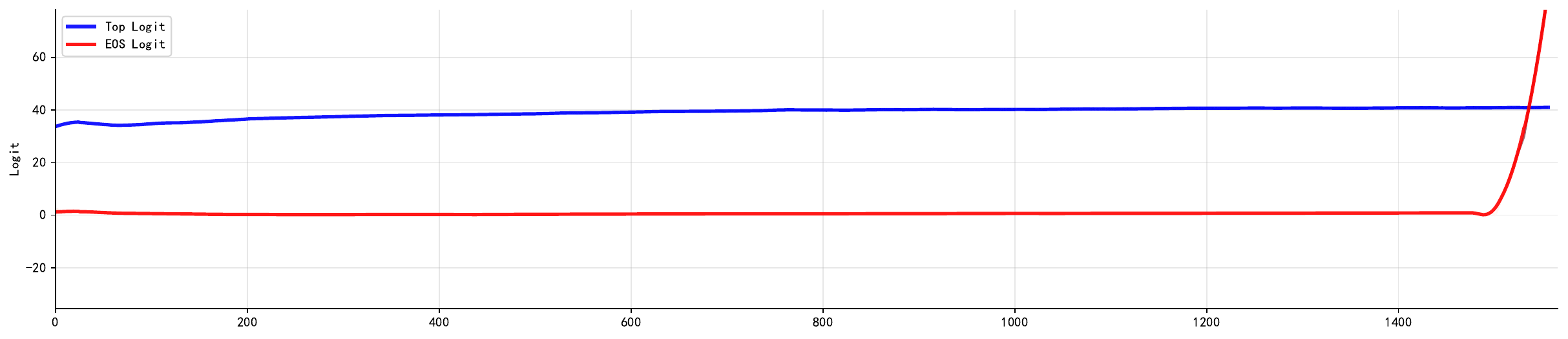}
    \caption{Effect of the EOS Indicator under End-Based Suppression ($\mathcal{L}_u = 1500$, $\eta = \frac{1}{32}$) on the P-DoS Attack with the Llama8B Model.}
    \label{fig:Llama-8b_pdos_1500_32.0}
\end{figure*}
\begin{figure*}[htbp]
    \centering
    \includegraphics[width=\textwidth]{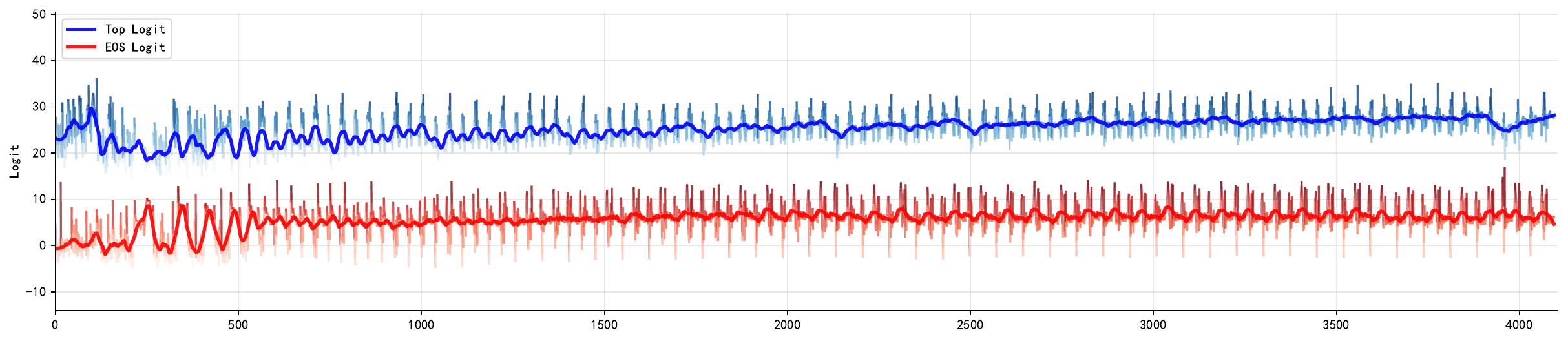}
    \caption{The eos indicator effect of executing GCG-DoS attack under the Llama8B model.}
    \label{fig:Llama-8b_GCG_Llama_4096_8.0}
\end{figure*}
\begin{figure*}[htbp]
    \centering
    \includegraphics[width=\textwidth]{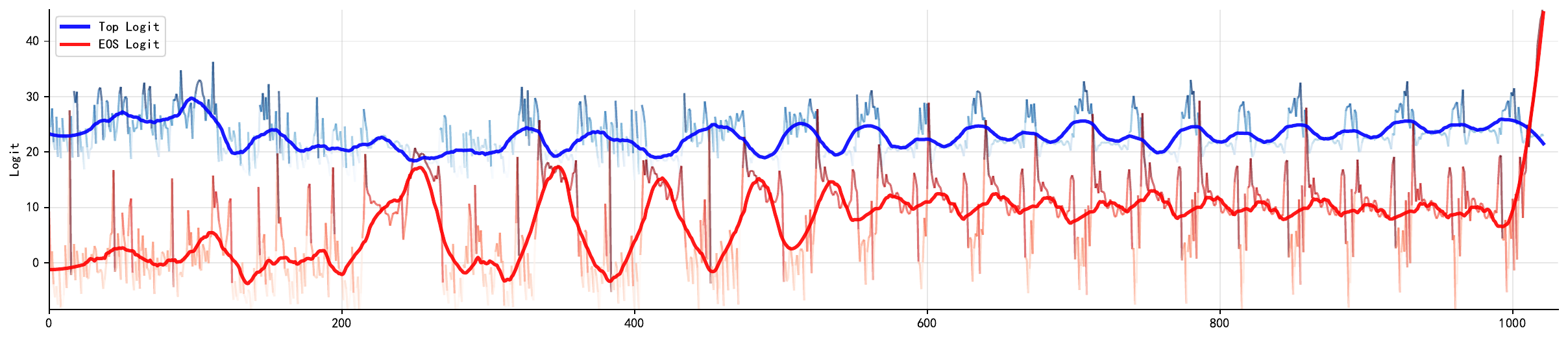}
    \caption{Effect of the EOS Indicator under End-Based Suppression ($\mathcal{L}_u = 1000$, $\eta = \frac{1}{8}$) on the GCG-DoS Attack with the Llama8B Model.}
    \label{fig:Llama-8b_GCG_Llama_1000_8.0}
\end{figure*}
\begin{figure*}[htbp]
    \centering
    \includegraphics[width=\textwidth]{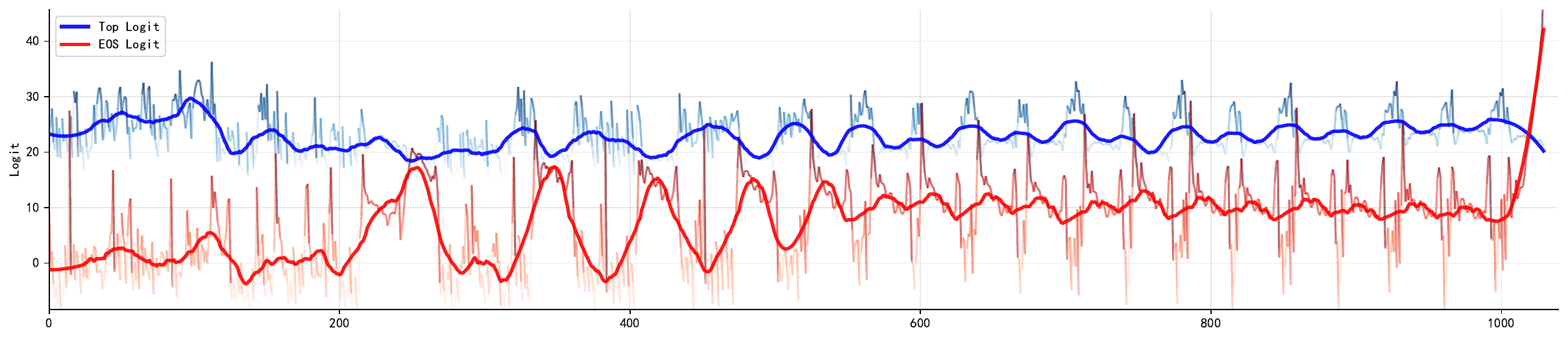}
    \caption{Effect of the EOS Indicator under End-Based Suppression ($\mathcal{L}_u = 1000$, $\eta = \frac{1}{16}$) on the GCG-DoS Attack with the Llama8B Model.}
    \label{fig:Llama-8b_GCG_Llama_1000_16.0}
\end{figure*}
\begin{figure*}[htbp]
    \centering
    \includegraphics[width=\textwidth]{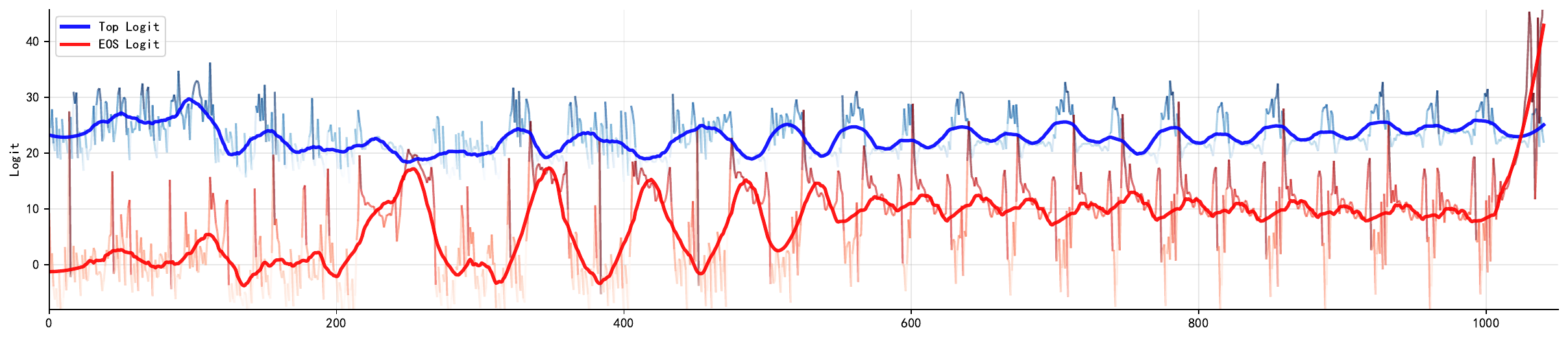}
    \caption{Effect of the EOS Indicator under End-Based Suppression ($\mathcal{L}_u = 1000$, $\eta = \frac{1}{24}$) on the GCG-DoS Attack with the Llama8B Model.}
    \label{fig:Llama-8b_GCG_Llama_1000_24.0}
\end{figure*}
\begin{figure*}[htbp]
    \centering
    \includegraphics[width=\textwidth]{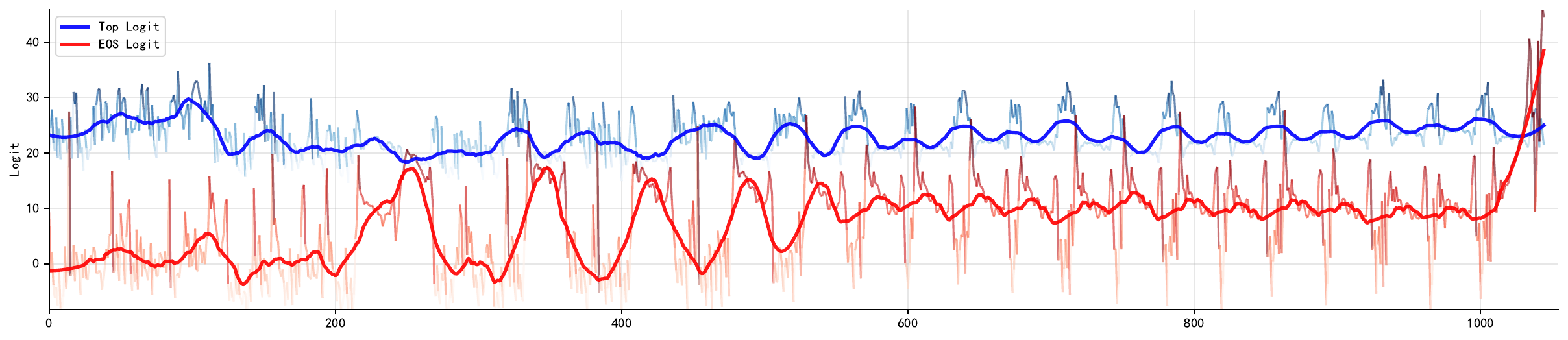}
    \caption{Effect of the EOS Indicator under End-Based Suppression ($\mathcal{L}_u = 1000$, $\eta = \frac{1}{32}$) on the GCG-DoS Attack with the Llama8B Model.}
    \label{fig:Llama-8b_GCG_Llama_1000_32.0}
\end{figure*}
\begin{figure*}[htbp]
    \centering
    \includegraphics[width=\textwidth]{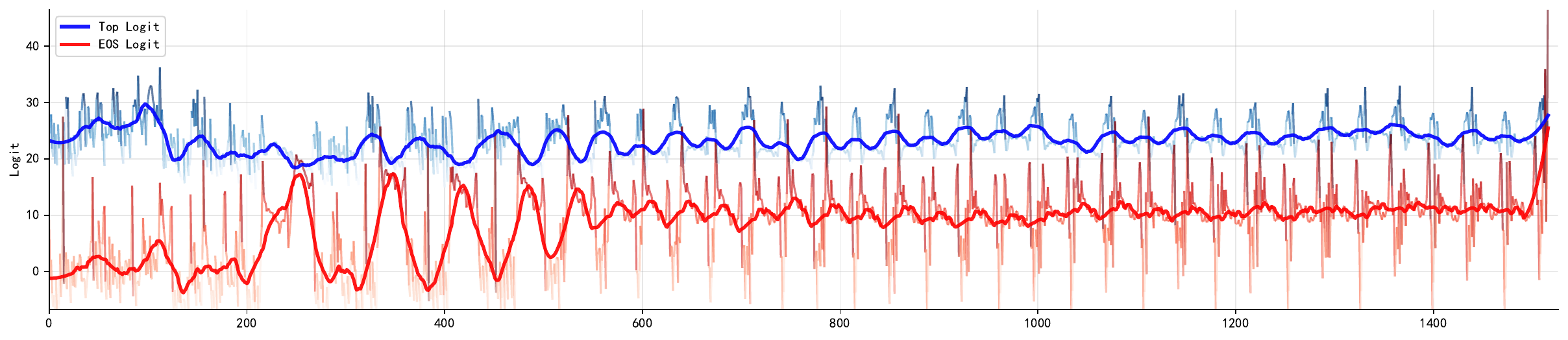}
    \caption{Effect of the EOS Indicator under End-Based Suppression ($\mathcal{L}_u = 1500$, $\eta = \frac{1}{8}$) on the GCG-DoS Attack with the Llama8B Model.}
    \label{fig:Llama-8b_GCG_Llama_1500_8.0}
\end{figure*}
\begin{figure*}[htbp]
    \centering
    \includegraphics[width=\textwidth]{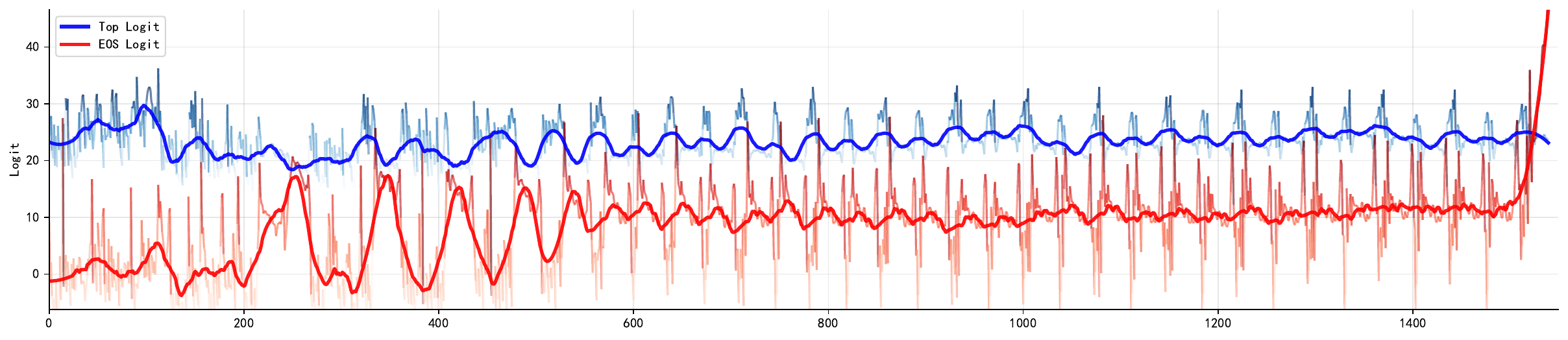}
    \caption{Effect of the EOS Indicator under End-Based Suppression ($\mathcal{L}_u = 1500$, $\eta = \frac{1}{16}$) on the GCG-DoS Attack with the Llama8B Model.}
    \label{fig:Llama-8b_GCG_Llama_1500_16.0}
\end{figure*}
\begin{figure*}[htbp]
    \centering
    \includegraphics[width=\textwidth]{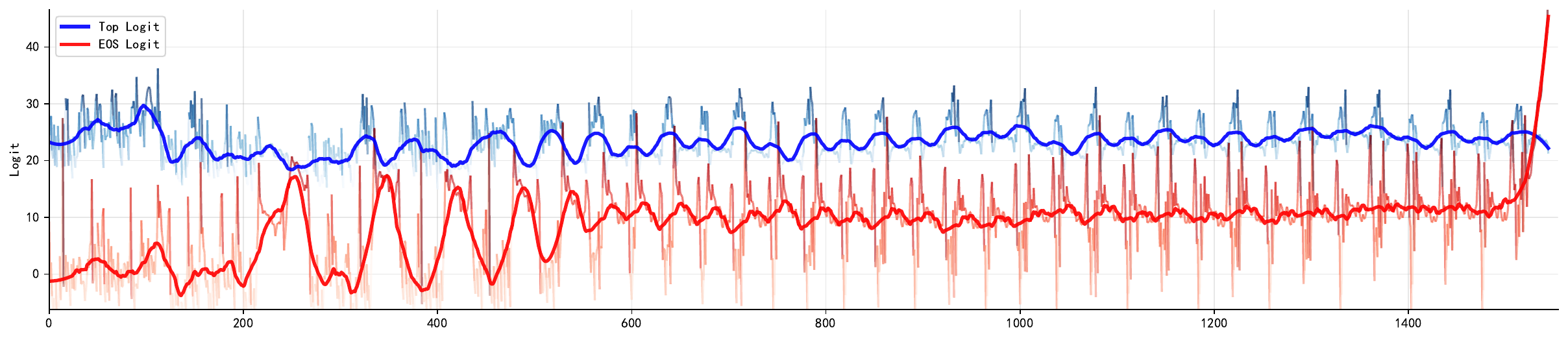}
    \caption{Effect of the EOS Indicator under End-Based Suppression ($\mathcal{L}_u = 1500$, $\eta = \frac{1}{24}$) on the GCG-DoS Attack with the Llama8B Model.}
    \label{fig:Llama-8b_GCG_Llama_1500_24.0}
\end{figure*}
\begin{figure*}[htbp]
    \centering
    \includegraphics[width=\textwidth]{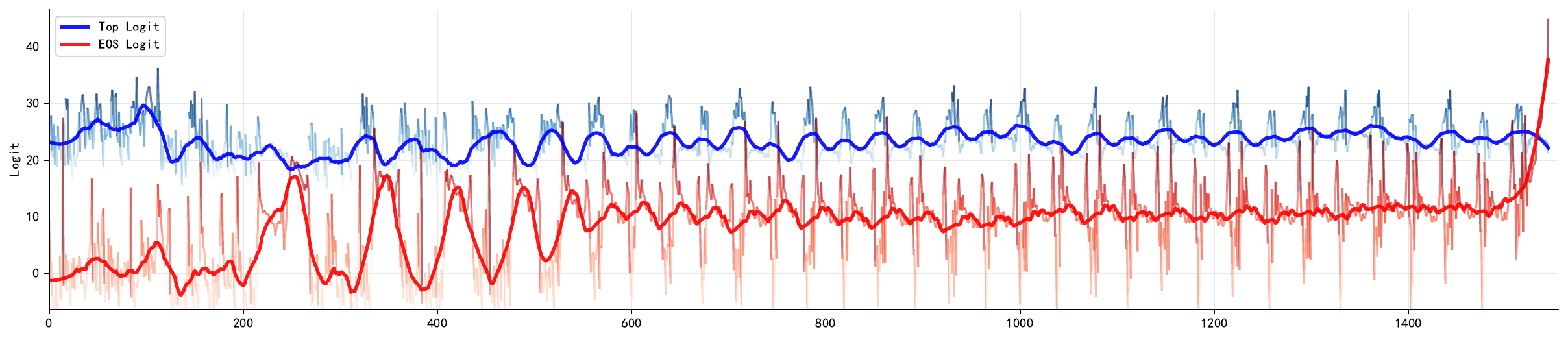}
    \caption{Effect of the EOS Indicator under End-Based Suppression ($\mathcal{L}_u = 1500$, $\eta = \frac{1}{32}$) on the GCG-DoS Attack with the Llama8B Model.}
    \label{fig:Llama-8b_GCG_Llama_1500_32.0}
\end{figure*}

\end{document}